\documentstyle[fleqn,12pt,epsfig]{article}
\newcommand{\be}{\begin{equation}}
\newcommand{\ee}{\end{equation}}

\newcommand{\eq}[1]{eq.~(\ref{#1})} 
\newcommand{\fig}[1]{Fig.~\ref{#1}}
\newcommand{\chii}{\chi_{{}_{\rm I}}}
\newcommand{\chir}{\chi_{{}_{\rm R}}}

\newcommand{\lessabout}{\raisebox{-.6ex}{\ $\stackrel{<}{\sim }$\ }}
\newcommand{\greaterabout}{\raisebox{-.6ex}{\ $\stackrel{>}{\sim }$\ }}
\begin{document}    
\setcounter{secnumdepth}{4}
\renewcommand\thepage{\ }
%
%
\begin{titlepage} 
%
\newcommand\reportnumber{501} 
\newcommand\mydate{September, 1998} 
\newlength{\nulogo} 
\settowidth{\nulogo}{\small\sf{hep-ph/9809403 }}
\title{\hfill\fbox{{\parbox{\nulogo}{\small\sf{
MADPH-98-1041\\
NUHEP \reportnumber\\ 
hep-ph/9809403 \\
\mydate
}}}}\vspace{0.5in} \\
{
Photon-Proton and Photon-Photon Scattering from 
Nucleon-Nucleon Forward Amplitudes
}}
 
\author{
M.~M.~Block
\thanks{Work partially supported by Department of Energy contract
DA-AC02-76-Er02289 Task B.}\vspace{-5pt}   \\
{\small\em Department of Physics and Astronomy,} \vspace{-5pt} \\ 
{\small\em Northwestern University, Evanston, IL 60208}\\
\vspace{-5pt}\\
%
E.~M.~Gregores\thanks{Work supported by Funda\c{c}\~ao de Amparo 
\`a Pesquisa do Estado de S\~ao Paulo (FAPESP).} 
and F.~Halzen
\thanks{Work partially supported by
Department of Energy Grant No.~DE-FG02-95ER40896 and
the University of Wisconsin Research
Committee with funds granted by the Wisconsin Alumni Research Foundation.}
\vspace{-5pt} \\
{\small\em Department of Physics,} \vspace{-5pt} \\
{\small\em University of Wisconsin, Madison, WI 53706}  \\
\vspace{-5pt}\\
%
G.~Pancheri
\vspace{-5pt} \\
{\small\em INFN-Laboratori Nazionali di Frascati,}\vspace{-5pt}  \\
{\small\em Frascati, Italy}\\
\vspace{-5pt}\\
%
}    
\vfill
\vspace{.5in}
\date{August 27, 1998}
\maketitle
\begin{abstract}

We show  that the data on $\gamma p$ and $\gamma\gamma$ interactions
can be derived from the $pp$ and $\bar p p$ forward scattering
amplitudes using vector meson dominance and the additive quark
model. The nucleon--nucleon data are parameterized using a model where
high energy cross sections rise with energy as a consequence of the
increasing numbers of soft partons populating the colliding
particles. We present detailed descriptions of the data on the total
and elastic cross sections, the ratio of the real to imaginary part of
the forward scattering amplitude, and on the slope of the differential
cross sections for $pp$, $\bar p p$, $\gamma p$, $\gamma\gamma$,
$\gamma p \rightarrow \gamma V$ and $\gamma\gamma
\rightarrow V_i V_j$ reactions, where $V= \rho, \omega, \phi$. We make
a wide range of predictions for future HERA and LHC experiments and
for $\gamma\gamma$ measurements at LEP.\
\end{abstract}
\end{titlepage} 
%
\pagenumbering{arabic}
\renewcommand{\thepage}{-- \arabic{page}\ --}  
%

\section{Introduction}

We show that the data on $\gamma p$ and $\gamma\gamma$ interactions
can be derived from the $pp$ and $\bar p p$ forward scattering
amplitudes using vector meson dominance (VMD) and the additive quark
model. We first show that the data on the total cross section, the
slope parameter $B$ and the ratio of the real to imaginary
part of the forward scattering amplitude $\rho$ for $pp$ and $\bar p p$
interactions, can be nicely described by a model
where high energy cross sections rise as a consequence of the
increasing numbers of soft partons populating the colliding
particles\cite{margolis}. The differential cross sections for the
Tevatron and LHC are predicted. Using this
parameterization of the hadronic  
forward amplitudes, we calculate the
photoproduction cross sections, slope and  
$\rho$ value from VMD and the additive quark model. We then obtain $\gamma\gamma$
cross  
sections which are again parameter-free,
demonstrating the approximate  
validity of the factorization theorem.

All cross sections will be computed in an eikonal formalism guaranteeing  
unitarity throughout:
\begin{eqnarray}
\sigma_{\rm tot}(s)&=&2\int\,\left[1-e^{-\chii (b,s)}\cos(\chir(b,s))\right]\,d^2\vec{b},
\end{eqnarray}
Here, $\chi $ is the complex eikonal ( $\chi=\chir + i\chii$) , and $b$ the impact 
parameter.
The even eikonal profile function $\chi^{\rm even}$ receives contributions from
quark-quark, quark-gluon and gluon-gluon interactions, and therefore
\begin{eqnarray}
\chi^{\rm even}(s,b) &=& \chi_{qq}(s,b)+\chi_{qg}(s,b)+\chi_{gg}(s,b)
\nonumber \\
&=& i\left [ \sigma_{qq}(s)W(b;\mu_{qq})
+ \sigma_{qg}(s)W(b;\sqrt{\mu_{qq}\mu_{gg}})
+ \sigma_{gg}(s)W(b;\mu_{gg})\right ]\, ,\label{chiintro}
\end{eqnarray}
where $\sigma_{ij}$ is the cross sections of the colliding partons, and
$W(b;\mu)$ is the overlap function in impact parameter space,
parameterized as the Fourier transform of a dipole form factor.  This
formalism is identical to the one used in ``mini-jet" models, as well
as in simulation programs for minimum-bias hadronic interactions such
as PYTHIA and SYBILL.

In this model hadrons asymptotically evolve into black disks of
partons. The rising cross section,
asymptotically associated with gluon-gluon interactions, is simply
parameterized by a normalization and energy scale, and two parameters:
$\mu_{gg}$ which describes the ``area" occupied by gluons in the
colliding hadrons, and $J (= 1+\epsilon)$. Here, $J$ is defined via
the gluonic structure function of the proton, which is assumed to
behave as $1/x^J$ for small x. It therefore controls the soft gluon
content of the proton, and it is meaningful that its value ($\epsilon
\simeq 0.05$) is consistent with the one observed in deep inelastic
scattering. The introduction of the quark-quark and quark-gluon
terms allows us to adequately parameterize the data at all energies,
since the ``size'' of quarks and gluons in the proton can be
different. We obtain $\mu_{qq}=0.89$ GeV, and $\mu_{gg}=0.73$
GeV. This indicates that gluons occupy a larger area of the proton
than do quarks.

The photoproduction cross sections are then calculated from this
parameterization of the hadronic forward amplitudes, assuming vector
meson dominance and the additive quark model. To this end , we introduce $P_{\rm had}$,
the 
probability that the photon interacts as a hadron. We
will use the value $P_{\rm had}=1/240$ which can be derived from
vector meson dominance. Our results show that its value is indeed
independent of energy. It is, however, uncertain by 30\% because it
depends on whether we relate photoproduction to $\pi$-nucleon or
nucleon-nucleon data (in other words, $\pi N$ and $NN$ elastic cross
sections only satisfy the additive quark model to this accuracy).
Subsequently, following reference \cite{fletcher}, we obtain $\gamma 
p$
cross sections from the assumption that, in the spirit of VMD, the
photon is a 2 quark state, in contrast with the proton which is a 3
quark state. Using the additive quark model and quark counting, the $\gamma p$ total 
cross section is obtained from the
even eikonal for $pp$ and $\bar p p$ by the substitutions:
\begin{eqnarray}
\sigma_{ij} &\rightarrow& \textstyle\frac{2}{3} \,\sigma_{ij}\; ,
\nonumber \\
\mu_i &\rightarrow& \sqrt{\textstyle\frac{3}{2}} \,\mu_i \, . 
\nonumber 
\end{eqnarray}

We will thus produce a parameter-free description of the total
photoproduction cross section, the phase of the forward scattering
amplitude and the forward slope for $\gamma p \rightarrow  Vp$,
where $V= \rho, \omega, \phi$. Interestingly, our results on the phase
of $Vp \rightarrow Vp$ are in complete agreement with the values
derived from Compton scattering results ($\gamma + p \rightarrow
\gamma + p$) using dispersion relations. We also calculate the total
elastic and differential cross sections for $\gamma p \rightarrow
Vp$. This wealth of data is accommodated without discrepancy.

The $\gamma\gamma$ cross sections are derived following the same  
procedure. We  now substitute
\begin{eqnarray}
\sigma_{ij} &\rightarrow& \textstyle\frac{4}{9} \,\sigma_{ij} \;,
\nonumber\\
\mu_i &\rightarrow& \textstyle\frac{3}{2}\,\mu_i \; , 
\nonumber
\end{eqnarray}
into the nucleon-nucleon even eikonal, and predict the total cross
section and differential cross sections for all reactions
$\gamma\gamma \rightarrow V_i V_j$  at a variety of energies, 
where $V= \rho, \omega, \phi$.

The high energy $\gamma\gamma$ total cross section \cite{exp} have 
been
measured by two experiments at LEP. These measurements yield new
information on its high energy behavior at center-of-mass energies in
excess of $\sqrt{s}=15$ GeV.  However, the two data sets
unfortunately disagree. We here point out that our analysis nicely
accommodates the L3 measurements. The analysis is sufficiently
restrictive to exclude the preliminary OPAL
results \cite{previous}. 
It is interesting to note that the small eikonal found in our model by fitting n-n 
data---as shown later---
enforces naturally the validity of the factorization theorem,
\begin{equation}
\frac{\sigma_{pp}}{\sigma_{\gamma p}}=\frac{\sigma_{\gamma p}}
{\sigma_{\gamma \gamma}},
\end{equation}
a result independent of the details of our model.
We find that the L3 data satisfy the factorization theorem
whereas the preliminary OPAL data do not. VMD and factorization are sufficient to
prevent one from adjusting $P_{\rm had}$, or any other parameters, to
change this conclusion.

An interesting theoretical issue emerges when it is noticed that we
applied the additive quark model to the full hadronic eikonal, not
just to the quark subprocesses in Eq.~2. This was not a choice---we found that the
data clearly enforced it.  For example, if we do not apply the quark
counting rules to the gluon-gluon subprocess, the model fails to
reproduce the forward slope of the $\gamma p \rightarrow \gamma V$
reactions, as well as the ratio of the imaginary to real part of the $\gamma p
\rightarrow \gamma p$ forward amplitude. It, in fact, fails 
dramatically.  This result may suggest either a static structure of
the nucleon where the gluons cluster around the original valence
quarks, as in the valon model \cite{valon}, or a dynamic picture in
which the gluons are associated with the interaction of the valence
quarks during the hadronic collision. This picture is reinforced by
correlation measurements between quarks and gluons, derived from the
observation of multiple parton final state in hadron collisions
\cite{eboli}.

We further emphasize the energy-independence of $P_{\rm had}$. In
other words, the shapes of the total cross sections for $\gamma p$ and
$\gamma \gamma$ reactions as a function of energy are completely fixed
by the shape of the nucleon-nucleon data.

\section{High Energy $\bar pp$ and $pp$ Scattering\label{pp}}
In this Section, we will discuss high energy $\bar p p$ and $pp$
scattering. In Section \ref{sec:fit}, we will discuss the theory of
our QCD-inspired eikonal and its implementation in fitting the
experimental data for $\sigma_{\rm tot},\ \rho$ and $B$ to determine
the parameters of the model.

In Section \ref{sec:ppelastic}, we will compare the experimental data
for the elastic scattering cross section, $\sigma_{\rm elastic}$ as a
function of energy with our predictions.

In Section \ref{sec:pppredictions}, we will show predictions for
differential elastic scattering at $\sqrt s=1800$ GeV, compared to
experimental data, and finally, a prediction for the differential
elastic scattering at the LHC.
\subsection{The Fit to High Energy $\bar pp$ and $pp$ Scattering Data
\label{sec:fit}}
We will fit all available high energy forward scattering data above 15
GeV, using {\em both} $\bar pp$ and $pp$, for
\begin{enumerate}
 \item $\sigma_{\rm tot}$, the total cross section,
\item $\rho$, the ratio of the real to the imaginary part of the forward 
scattering amplitude,
\item $B$, the logarithmic slope of the differential elastic scattering 
cross section in the forward direction. 
\end{enumerate}
We insist that our QCD-inspired model satisfies:
\begin{enumerate}
\item crossing symmetry, {\em i.e.,} be either even or odd under the 
transformation  $E\rightarrow -E$, where E is the laboratory energy. This 
 allows us to 
simultaneously describe $\bar pp$ and $pp$ scattering.
\item unitarity.  We will use an eikonal formalism to guarantee this.
\item analyticity.  We need this to calculate the {\em phase} of the 
forward scattering amplitude and hence, the $\rho$ value.
\item the Froissart bound. Asymptotically, we expect that the total 
cross section will rise as $(\log s)^2$. 
\end{enumerate}
The formalism needed to derive $\sigma_{\rm tot}$, $\rho$ and $B$ from
an eikonal are given in Appendix \ref{app:eikonal}, in sections
\ref{app:sigtot}, \ref{app:rho} and \ref{app:B}, in
eqns. (\ref{sigtot}), (\ref{rho}) and (\ref{Bfinal}),
respectively. Details on the analyticity are given in Ap pendices
\ref{app:evencontribution} and \ref{app:oddeikonal}. The even portion
(under crossing) of our QCD-inspired eikonal, as noted in
\eq{chiintro}, contains quark-quark, quark-glue and glue-glue
components, of which the glue-glue portion dominates at high energy. A
detailed parameterization of this portion is given in Appendix
\ref{app:QCDeven}, sections \ref{app:siggg} and
\ref{app:sigggevaluation}. The even eikonal we finally use is given in
\eq{evenanalytic}.

We show in Appendix \ref{app:highenergysigmagg} that the Froissart
bound is satisfied by our glue-glue interaction, and that,
asymptotically, the cross section is given by
\begin{equation}
\sigma_{tot}=2\pi \left(\frac{\epsilon}{\mu_{gg}}\right)^2\log^2\frac{s}{s_0},
\end{equation}
as seen in \eq{Froissart}. The parameter $\epsilon$ is defined via the
gluonic structure of the proton, which is assumed to behave as
$1/x^{1+\epsilon}$, for small $x$. The mass $\mu_{gg}$ describes the
area occupied by the gluons in the colliding hadrons. Both are fitted
from experiment, with $\epsilon\approx 0.05$ and $\mu_{gg}\approx
0.73$ GeV. These two parameters, along with the threshold mass
$m_0\approx 0.6$ GeV and the strength of the glue-glue interaction,
$C_{gg}$, are all that is required to specify the glue-glue
interaction. These 4 parameters dominate the high energy behavior of
the nucleon-nucleon cross section and are the critical elements of our
fit.

The quark-quark and quark-glue portions are discussed in Appendices
\ref{app:sigqq} and \ref{app:sigqg}, and are simulated by a constant
strength $C$, a Regge descending trajectory strength $C_{\rm Regge\
even}$, a strength $C_{\rm qg\ log}$ for a log term, an energy scale
$\sqrt s_0$ for the log term and a quark size $\mu_{qq}$, as detailed
in \eq{finaleven} in Appendix \ref{app:evencontribution}.  We show how
to make the even eikonal analytic in section
\ref{app:evencontribution}. Details of the odd eikonal are given in
Appendix \ref{app:oddeikonal}, and the odd eikonal, which contains two
parameters, a strength $C_{\rm odd}$ and a size $\mu_{\rm odd}$, is
given in \eq{oddanalytic}.

In all, 11 parameters are used in the theoretical model. The low
energy region, for $\sqrt s\lessabout 25$ GeV, where the {\em
differences} between $\bar p p$ and $pp$ scattering are substantial,
largely determine the 7 parameters necessary to fit the odd eikonal
and the quark-quark and the quark-glue portions of the even
eikonal. Thus, they largely decouple from the high energy behavior,
which depends on {\em only} 4 of these quantities. Hence, for $\sqrt
s\greaterabout 25$ GeV, where there is little difference between 
$\bar p p$ and $pp$ scattering, we really need only 4 parameters for our
model.

\begin{figure}[htbp]
\begin{center}
\mbox{\epsfig{file=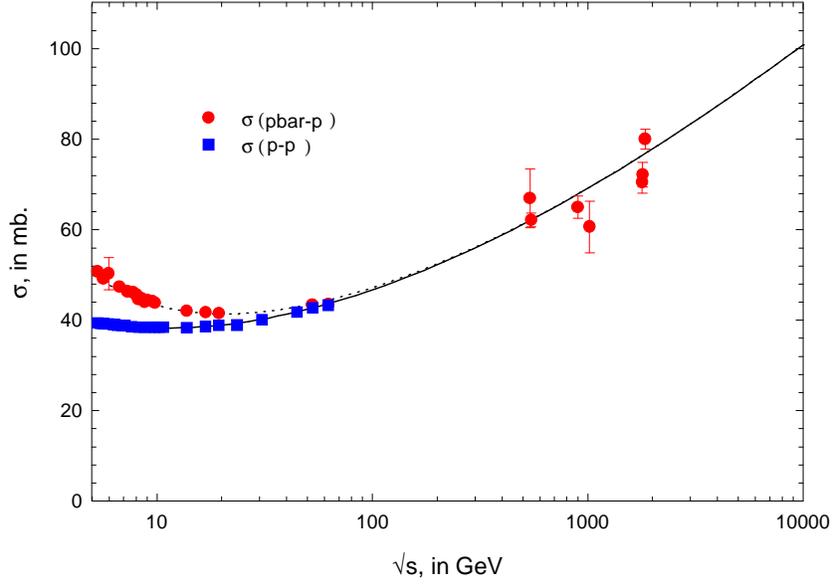,width=4.5in,%
bbllx=100pt,bblly=350pt,bburx=537pt,bbury=660pt,clip=}}
\end{center}
\caption[]{\footnotesize The total cross section $\sigma_{\rm tot}$, in 
mb {\em vs.} $\sqrt s$, in GeV, for $pp$ and $\bar p p$  scattering.  
The solid line and squares are for $pp$ and the dotted line and
circles are for $\bar p p$.}
\label{fig:1sigtot}
\end{figure}

\begin{figure}[htp]
\begin{center}
\mbox{\epsfig{file=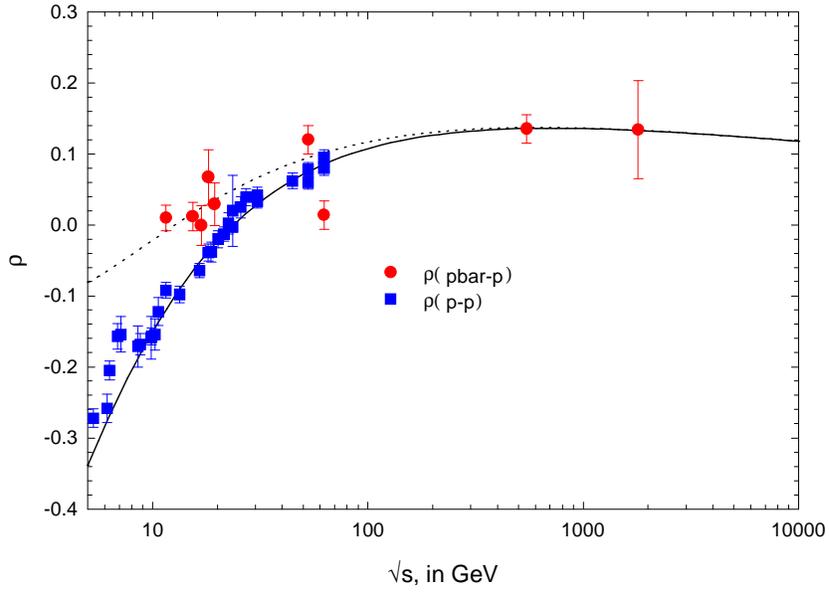,width=4.5in,%
bbllx=100pt,bblly=358pt,bburx=537pt,bbury=660pt,clip=}}
\end{center}
\caption[]{\footnotesize The ratio of the real to imaginary part of the 
forward scattering amplitude, $\rho$ {\em vs.} $\sqrt s$, in GeV, for 
$pp$ and $\bar p p$ scattering.
 The solid line and squares are for $pp$  and the dotted 
line and circles are for $\bar p p$.}
\label{fig:1rho}
\end{figure}

\begin{figure}[htbp] 
\begin{center}
\mbox{\epsfig{file=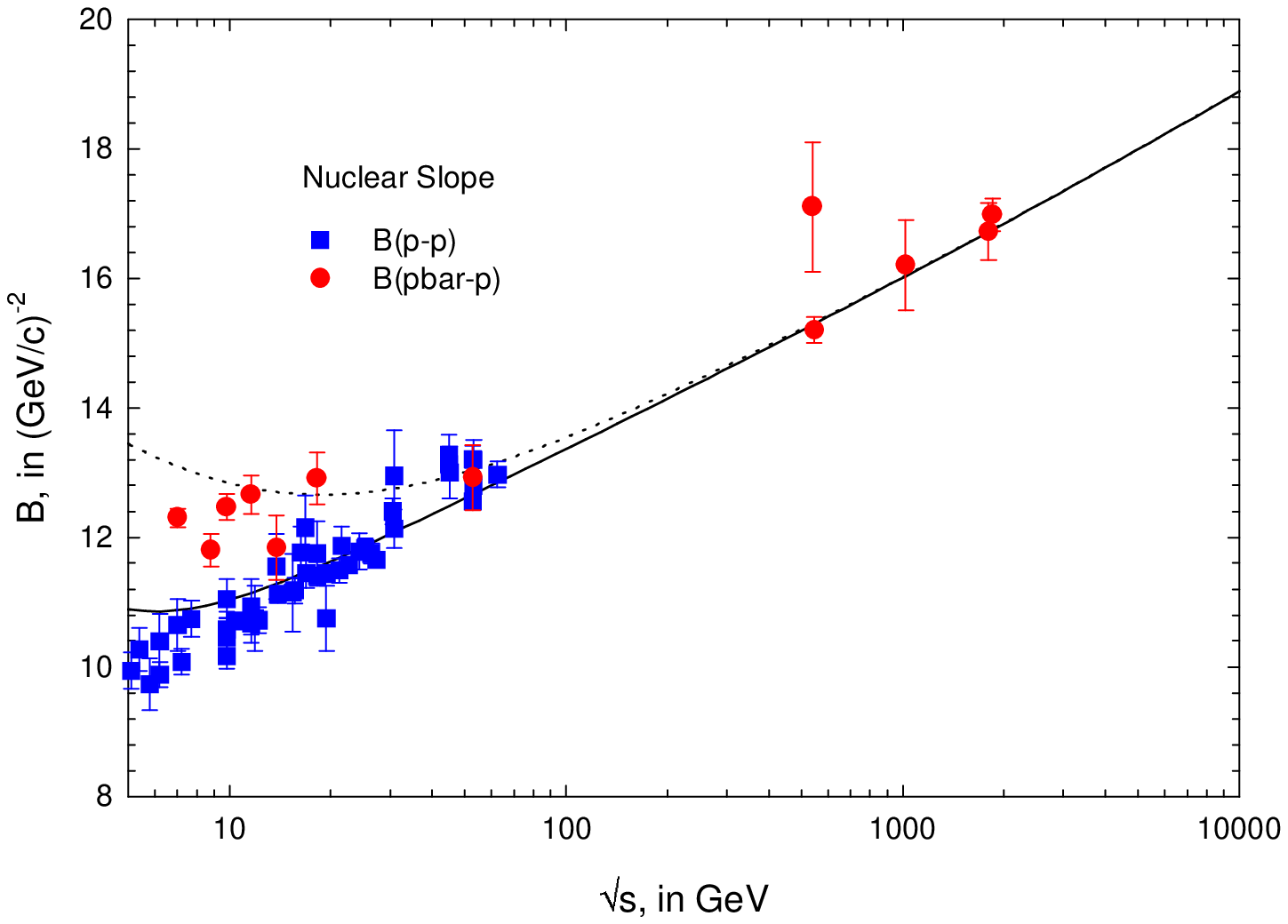,width=4.5in,%
bbllx=100pt,bblly=358pt,bburx=537pt,bbury=660pt,clip=}}
\end{center}
\caption[]{\footnotesize The nuclear slope parameter B, in (GeV/c)$^{-2}$ 
{\em vs.} $\sqrt s$, in GeV, for elastic $pp$ and $p\bar p$ scattering.  
The solid line and squares are for $pp$ and the dotted line and
circles are for $\bar p p$.}
\label{fig:1b}
\end{figure}

 We use in the fit {\em all} of the highest energy cross sections
 available, (E710\cite{E710}, CDF\cite{CDF} and the unpublished
 Tevatron value\cite{newpp}) which anchor the upper end of our cross
 section curves.  The results of the fit are shown in
 Figs. \ref{fig:1sigtot}, \ref{fig:1rho} and \ref{fig:1b}. The total
 cross sections $\sigma_{\rm tot}$ for $\bar p p$ (dotted line and
 circles) and for $pp$ (solid line and squares) are plotted against
 $\sqrt s$, the cms energy, in \fig{fig:1sigtot}.
The $\rho$ values (the ratio of the real to the imaginary portion of
the forward scattering amplitude) are plotted in \fig{fig:1rho} using
the same conventions, 
%
and the nuclear slope B values are similarly plotted in \fig{fig:1b}.

It can be seen from these figures that we have a quite satisfactory
description of all 3 quantities, for both $\bar pp$ and $pp$
scattering. The $\chi^2$ of the fit is reasonably good (considering
the large spreads in the experimental data---in $B$, in particular, as
well as the discrepancies in t he highest energy cross sections),
giving a $\chi^2$ of 130.3, where 75 was expected. The cross section
fit of \fig{fig:1sigtot} splits the difference between the values at
$\sqrt s=1800$ GeV. From \fig{fig:1rho}, we note that the fit to
$\rho$ is anchored at $\sqrt s=550$ GeV by the very accurate
measurement\cite{ua42} of UA4/2 and passes through the E710
point\cite{E710rho}.
 
The statistical uncertainty of the fitted parameters is such that at
25 GeV, the cross section predictions are statistically uncertain to
$\approx 1.3$\%, at 500 GeV are uncertain to $\approx 1.6$\% and at
2000 GeV are uncertain to $\approx2.5$\% .
\subsection{Predictions for Elastic Scattering Cross Sections\label{sec:ppelastic}}

\begin{figure}[htbp] 
\begin{center}
\mbox{\epsfig{file=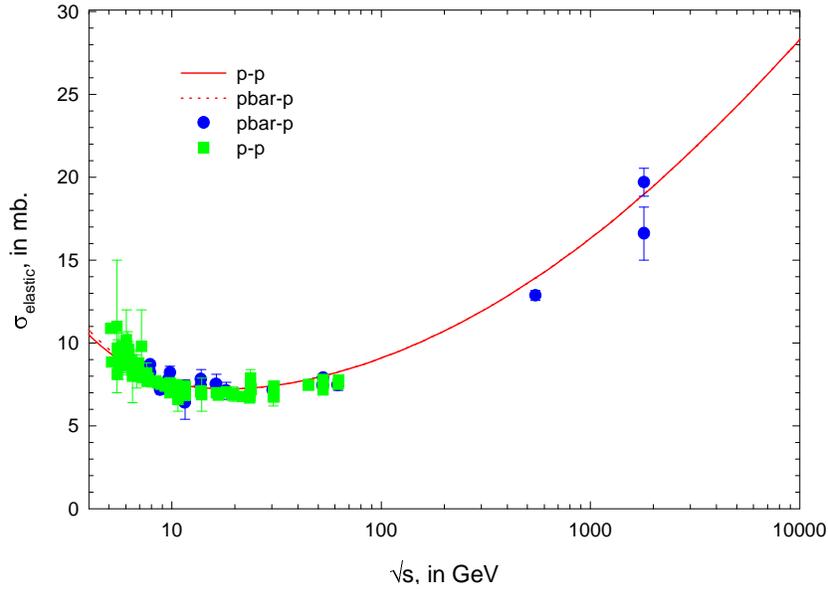,width=4.5in,%
bbllx=100pt,bblly=358pt,bburx=537pt,bbury=660pt,clip=}}
\end{center}
\caption[]{\footnotesize Elastic scattering cross sections, 
$\sigma_{\rm elastic}$, in mb {\em vs.} $\sqrt s$, in GeV, 
for $pp$ and $p\bar p$ scattering. The solid line and squares 
are for $pp$  and the dotted 
line and circles are for $\bar p p$.}
\label{fig:1sigel}
\end{figure}

We now have all the parameters needed to specify our eikonal.  In
\fig{fig:1sigel} we have plotted our prediction for the elastic cross
section $\sigma_{\rm elastic}$, in mb {\em vs.} the cms energy $\sqrt
s$, in GeV, along with the available data for both $\bar pp$ and $pp$.
The agreement is excellent.
%

We note that $\sigma_{\rm elastic}$ is rising more sharply with energy
than does the total cross section $\sigma_{\rm to}$. Comparing
\fig{fig:1sigtot} with \fig{fig:1sigel}, we see that the ratio of the
elastic to total cross section is rising significantly with energy.
The ratio is, of course, bounded by the value for the black
disk\cite{bc,bcasymptopia}, {\em i.e.,} 0.5, as the energy goes to
infinity.
\subsection{Predictions for $\bar p p$ Elastic Differential Scattering 
Cross Sections at 1800 GeV\label{sec:pppredictions}}

\begin{figure}[htbp] 
\begin{center}
\mbox{\epsfig{file=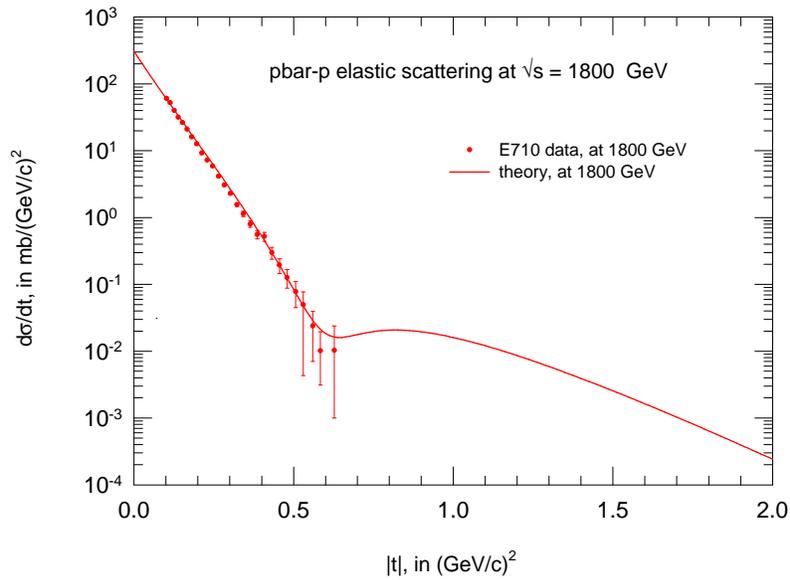,width=4.5in,%
bbllx=90pt,bblly=250pt,bburx=580pt,bbury=610pt,clip=}}
\end{center}
\caption[]{\footnotesize The elastic differential scattering cross section 
$\frac{d\sigma}{dt}$, in mb/(GeV/c)$^2$ {\em vs.} $|t|$, in (GeV/c)$^2$, 
for the reaction $\bar pp\rightarrow\bar pp$ at $\sqrt s=1800$ GeV. 
The solid curve is the prediction at the Tevatron Collider and the data 
points are from E710.}
\label{fig:1ds1800}
\end{figure}

>From \eq{dsdt}, we can now calculate $\frac{ds}{dt}$, the elastic
differential cross section as a function of $|t|$, for various values
of $\sqrt s$.  The calculated differential cross section at the
Tevatron ($\sqrt s=1800$ GeV) is shown in \fig{fig:1ds1800} and
compared with the experimental data from E710\cite{E710slope}.  The
agreement over 4 decades is striking.
\subsection{Predictions for $\bar p p$ at the LHC \label{sec:pppredictionsLHC}}
With the parameters we obtained from our fit, we predict the total
cross section at the LHC (14 TeV) as $\sigma_{\rm tot}=108.0\pm 3.4$
mb, where the error is due to the statistical errors of the fitting
parameters.

\begin{figure}[htbp]
\begin{center}
\mbox{\epsfig{file=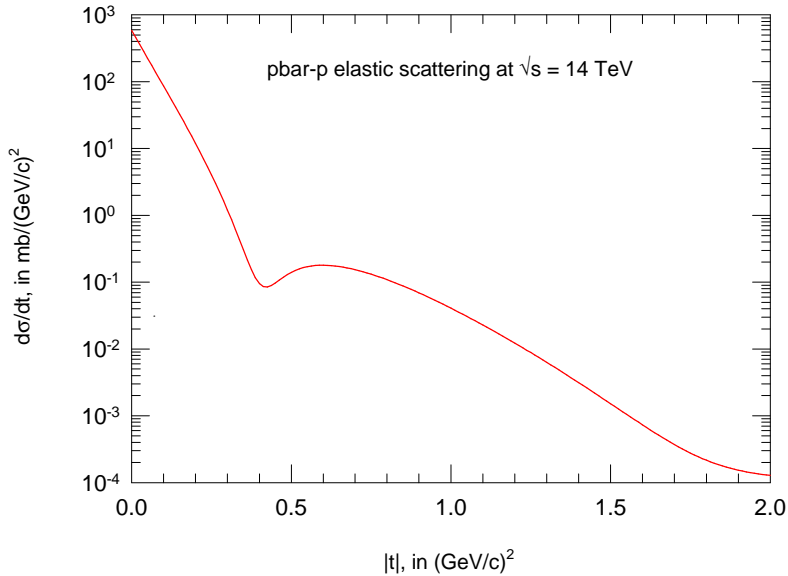,width=4.5in,%
bbllx=90pt,bblly=250pt,bburx=580pt,bbury=610pt,clip=}}
\end{center}
\caption[]{\footnotesize The predicted elastic differential scattering 
cross section $\frac{d\sigma}{dt}$, in mb/(GeV/c)$^2$ {\em vs.} $|t|$, 
in (GeV/c)$^2$, for the reaction $\bar pp\rightarrow\bar pp$ at 
$\sqrt s=14$ TeV, at the LHC.}
\label{fig:1ds14000}
\end{figure}

Our prediction for the differential cross section for $\sqrt s=14$
TeV, the energy of the LHC, is plotted in \fig{fig:1ds14000}.
%
%
It will be a challenge to the LHC to measure this cross section and to
confirm the predicted structure in $|t|$. In particular, at small
$|t|$, we predict that the curvature parameter $C$ (see Appendix
\ref{app:curvature} for details) is {\em negative}.  For energies much
lower than 1800 GeV, the observed curvature has been measured as {\em
positive}. For 1800 GeV, we see from \fig{fig:1ds1800} that the
curvature parameter $C$ is compatible with being zero.  Block and
Cahn\cite{bc,bcasymptopia} had pointed out that they expected the
curvature to go through zero near the Tevatron energy an d that it
should become negative thereafter. Basically, the argument is that
experimentally, the curvature for the then available energies was
positive. However, it was expected that the scattering asymptotically
would approach that of a sharp disk. The curvature of a (gray) sharp
disk\cite{bc,bcasymptopia} is always negative, $C=-R^4/192$, where
$R$ is the radius of the disk. Thus, the curvature had to pass through
zero as the energy increased.  They called `asymptopia' the energy
region (energies much larger than the Tevatron Collider) where the
scattering approached that of a sharp disk .
%
%
%
\section{$\gamma p$ Reactions\label{sec:gammap}}
In our model, when the photon interacts strongly, it behaves like a
two quark system.  Taking the quark model literally, the eikonal for
$\gamma p$ scattering is obtained by rewriting the even eikonal with
the substitutions
\begin{eqnarray}
\sigma_{ij} &\rightarrow& \textstyle\frac{2}{3} \,\sigma_{ij}\nonumber
\; ,\\
\mu_i &\rightarrow& \sqrt{\textstyle\frac{3}{2}} \,\mu_i \label{2/3} \, ,
\end{eqnarray}
as {\samepage{\begin{eqnarray}
\chi^{\gamma p}(s,b) 
&=& i\left [\frac{2}{3}
\sigma_{qq}(s)W\left(b;\sqrt{\frac{3}{2}}\mu_{qq}\right) + \frac{2}{3}
\sigma_{qg}(s)W\left(b;\sqrt{\frac{3}{2}}\sqrt{\mu_{qq}\mu_{gg}}\right)
\right.\nonumber\\
&&\qquad \left. + \frac{2}{3}
\sigma_{gg}(s)W\left(b;\sqrt{\frac{3}{2}}\mu_{gg}\right)\right ]\,
.\label{chigammap}
\end{eqnarray}
}}
\subsection{$\gamma p$ Total Cross Section Prediction}
Using vector dominance and the $\gamma p$ eikonal of \eq{chigammap},
we can now write, using \eq{sigtot},
\begin{eqnarray}
\sigma_{\rm tot}^{\gamma p}(s)
=&2P_{\rm had}\int\,\left[1-e^{-{\chii}^{\gamma p}
(b,s)}\cos({\chir}^{\gamma p}
(b,s))\right]\,d^2\vec{b},\label{sigtotgammap}
\end{eqnarray}
where $P_{\rm had}$ is the probability that a photon will interact as
a hadron. We will use the value $P_{\rm had}=1/240$, which is found by
fitting the low energy $\gamma p$ data.  This value is very close to
that derived from VDM. Using (see Table XXXV, p.393 of
ref. \cite{bauer}) $\frac{f_{\rho}^2}{4\pi}=2.2$,
$\frac{f_{\omega}^2}{4\pi}=23.6$ and $\frac{f_{\phi}^2}{4\pi}=18.4$,
we find $\Sigma_{V}\frac{4\pi\alpha}{f_V^2}=1/249$, where
$V=\rho,\omega\phi$.

\begin{figure}[htbp] 
\begin{center}
\mbox{\epsfig{file=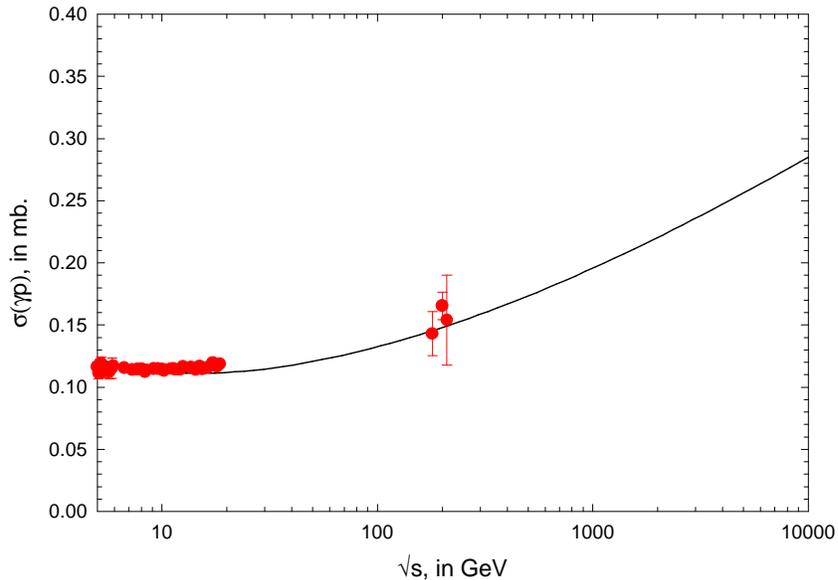,width=4.5in,%
bbllx=100pt,bblly=358pt,bburx=537pt,bbury=660pt,clip=}}
\end{center}
\caption[]{\footnotesize The total cross section, $\sigma_{\rm tot}$, in 
mb {\em vs.} $\sqrt s$, in GeV,  for $\gamma p$ scattering. The solid 
curve is the predicted total cross section.}
\label{fig:2sigtot}
\end{figure}

With all eikonal parameters fixed from our nucleon-nucleon fits and
our choice of $P_{\rm had}=1/240$, we can now calculate $\sigma_{\rm
tot}^{\gamma p}(s)$. Our prediction is given in \fig{fig:2sigtot},
%
where $\sigma_{\rm tot}^{\gamma p}(s)$ is plotted against the cms
energy. The fit reproduces the rising cross section for $\gamma p$,
using parameters fixed by nucleon-nucleon scattering. We comment here
that this prediction only uses the 9 parameters of the even eikonal,
of which but 4 are important in the upper energy region. The accuracy
of our predictions are $\approx 1.5\%$, from the statistical
uncertainty in our eikonal parameters.
\subsection{`Elastic' $\gamma p$ Scattering\label{sec:elasticgp}} 
We consider as `elastic' scattering the three vector reactions
\begin{eqnarray}
\gamma+p&\rightarrow &\rho_{\rm virtual}+p\rightarrow\rho + p \nonumber\\
\gamma+p&\rightarrow &\omega_{\rm virtual}+p\rightarrow\omega + p \nonumber\\
\gamma+p&\rightarrow &\phi_{\rm virtual}+p\rightarrow\phi + p ,\label{Vpreactions}
\end{eqnarray}
where the photon virtually materializes as a vector meson, which then
elastically scatters off of the proton. The strengths of these
reactions is $\approx \alpha$, the fine-structure constant, times a
strong interaction cross section. The true elastic cross section is
given by Compton scattering on the proton, $\gamma
+p\rightarrow\gamma+p$, which we can visualize as
\begin{eqnarray}
\gamma+p&\rightarrow &\rho_{\rm virtual}+p\rightarrow\rho + p 
\rightarrow \gamma+p\nonumber\\
\gamma+p&\rightarrow &\omega_{\rm virtual}+p\rightarrow\omega + p
\rightarrow \gamma+p\nonumber\\
\gamma+p&\rightarrow &\phi_{\rm virtual}+p\rightarrow\phi + p
\rightarrow \gamma+p,\label{Compton}
\end{eqnarray}
is clearly $\alpha^2$ times a strong interaction cross section, and
hence is much smaller than `elastic' scattering of \eq{Vpreactions}.
Thus, we justify using \eq{sigtotgammap} to calculate the {\em total}
cross section which we compare with experiment, since only reactions
with a photon in the final state (down by $\approx\alpha$) are
neglected.
\subsubsection{Predictions of $\rho$ and $B$ for `Elastic' $\gamma p$ 
Reactions\label{sec:rho&Bgp}}
Using the philosophy of `elastic' scattering expressed in
\eq{Vpreactions}, and eqns. (\ref{rho}) and (\ref{Bfinal}), we can
immediately write
\begin{eqnarray}
\rho(s)
&=&\frac{{\rm Re}\left\{i(\int 1-e^{i\chi^{\gamma
p}(b,s)})\,d^2\vec{b}\right\}} {{\rm Im}\left\{i(\int
(1-e^{i\chi^{\gamma p}(b,s)})\,d^2\vec{b}\right\}},\label{rhogp}
\end{eqnarray}
and
\be
B=\frac{1}{2}\frac{\int\left(1-e^{i\chi^{\gamma
p}(b,s)}\right)\,b^2\,\,d^2\vec{b}} {\int\left(1-e^{i\chi^{\gamma
p}(b,s)}\right)\,\,d^2\vec{b}}.\label{Bgp}
\ee
We see from eqns. (\ref{rhogp}) and (\ref{Bgp}) that the predictions
for both $\rho$ and the slope $B$ are free of any $P_{\rm had}$
factors and hence are {\rm independent} of normalization---thus being
the same for either $\rho p,\ \omega p$ or $\phi p$ final states.

\begin{figure}[htbp] 
\begin{center}
\mbox{\epsfig{file=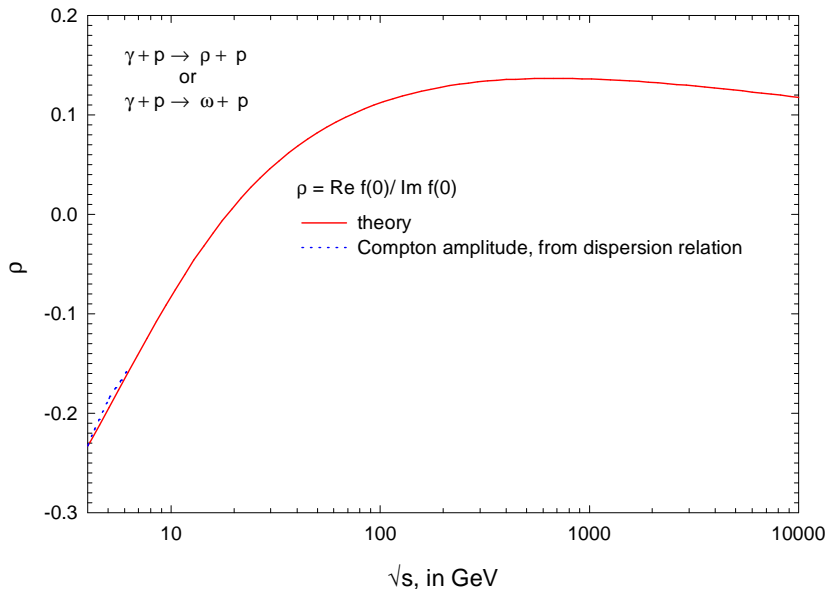,width=4.5in,%
bbllx=100pt,bblly=358pt,bburx=537pt,bbury=660pt,clip=}}
\end{center}
\caption[]{\footnotesize The solid curve is the predicted ratio of the 
real to imaginary part of the forward scattering amplitude for the
`elastic' reactions , $\gamma +p\rightarrow V_i + p$ scattering
amplitude, where $V_i$ is $\rho^0$, $\omega^0$ or $\phi^0$ {\em vs.}
$\sqrt s$, in GeV. The dotted curve is ratio of the real to imaginary
part of the forward scattering amplitude for Compton
scattering{\protect{\cite{gilman}}}, $\gamma +p\rightarrow\gamma + p$,
found from dispersion relations.  It has been slightly displaced from
the solid curve for clarity in viewing.}
\label{fig:2rhovec}
\end{figure}

In \fig{fig:2rhovec} the value of $\rho$ from \eq{rhogp} is the solid
curve, plotted as a function of $\sqrt s$. Damashek and
Gilman\cite{gilman} calculated the $\rho$ value for Compton scattering
on the proton, using dispersion relations, {\em i.e.,} the {\em true}
elastic scattering reaction for photon-proton scattering.  We compare
this calculation, the dotted line in \fig{fig:2rhovec}, with our
prediction of $\rho$ (the solid line).  The agreement is so close that
we had to move the two curves apart so that they may be viewed more
clearly---this gives us confidence in our approach.
%

\begin{figure}[htbp]
\begin{center}
\mbox{\epsfig{file=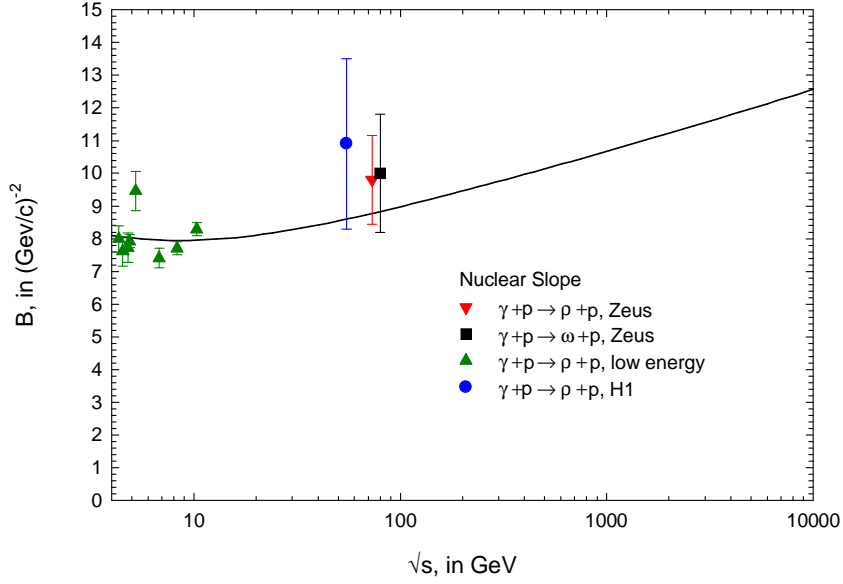,width=4.5in,%
bbllx=100pt,bblly=400pt,bburx=480pt,bbury=670pt,clip=}}
\end{center}
\caption[]{\footnotesize The predicted nuclear slope parameter B, 
in (GeV/c)$^{-2}$ {\em vs.} $\sqrt s$, in GeV, for the `elastic'
reaction $\gamma +p\rightarrow V_i + p$, where $V_i$ is $\rho^0$,
$\omega$ or $\phi$ scattering. The solid curve is the prediction.  For
the reaction $\gamma + p\rightarrow\rho^0 + p$, the inverted
triangles are the Zeus data, the circles are the H1 data and the
triangles are the low energy data.  For the reaction $\gamma +
p\rightarrow\omega + p$, the squares are the Zeus data.}
\label{fig:2bvec}
\end{figure}

In \fig{fig:2bvec} the solid curve is our prediction for the slope $B$
against the energy $\sqrt s$.  The available experimental data for
`elastic' $\rho p$ and $\omega p$ final states are also
plotted. Again, the agreement of theory and experiment is quite
satisfactory.

We note that the predictions of $\rho$ and $B$ are very critical to
our analysis.  We have assumed that in some manner, the gluons are
``attached'' to the quarks---when we have a two quark system, such as
the photon, the factors of 2/3 times a cross section and $\sqrt {3/2}$
times a $\mu$ of \eq{2/3} in the even eikonal of \eq{chigammap} are
the {\em same} for glue-glue as for quark-quark. If we relax this
assumption, and only use these factors in the quark, then we get sharp
disagreement with our predicted $\rho$ value---being considerably
larger than the Compton value. This is further exacerbated in the
predictions for $B$, with slopes from 11 (at 5 GeV) to 16
(GeV/c)$^{-2}$ (at 80 GeV), which are much larger than the
experimental values. We stress that these conclusions are {\em
independent} of our choice of $P_{\rm had}$.  Thus, our model clearly
has dynamical consequences whi ch will be discussed in detail later.
\subsubsection{Predictions of $\sigma_{\rm elastic}$ and $\frac{d\sigma}{dt}$ 
for `Elastic' $\gamma p$ Reactions\label{sec:dsdt}}
To find the elastic cross sections $\sigma_{\rm elastic}^{Vp}$ and
differential cross sections $d\sigma^{Vp}/dt$ as a function of energy,
using \eq{sigel}, we write
\begin{eqnarray}
\sigma_{\rm elastic}^{Vp}(s)
&=&P_{\rm had}^{Vp}\int\left|1-e^{i\chi^{\gamma
p}(b,s)}\right|^2\,d^2\vec{b},\label{sigelgp}
\end{eqnarray}
where $P_{\rm had}^{Vp}$ is the appropriate probability for a photon
to turn into $V$, where $V=\rho,\ \omega$ or $\phi$, respectively.

Similarly, the differential scattering cross section is, using
\eq{dsdt2}, given by
\be
\frac{d\sigma^{Vp}}{dt}(s,t)=\frac{P_{\rm had}^{Vp}}{4\pi}
\left|\int J_0(qb)(1-e^{i\chi^{\gamma p}(b,s)})\,d^2\vec{b}\,\right|^2,
\label{dsdt2gp}
\ee
where $t=-q^2$. The same factors $P_{\rm had}^{Vp}$ are used in
\eq{dsdt2gp} as in \eq{sigelgp}.

\begin{figure}[htbp] 
\begin{center}
\mbox{\hspace{0.2in}\epsfig{file=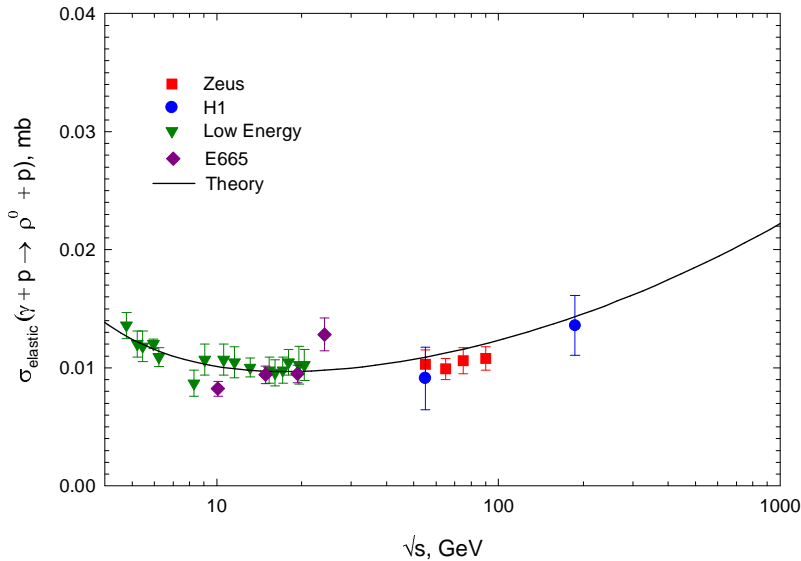,width=4.5in,%
bbllx=80pt,bblly=330pt,bburx=540pt,bbury=650pt,clip=}}
\end{center}
\caption[]{\footnotesize The `elastic' photoproduction cross section, 
$\sigma_{\rm elastic}$, in mb {\em vs.} $\sqrt s$, in GeV, for the
reaction $\gamma +p\rightarrow\rho^0 + p$. The solid curve is the
predicted cross section, the squares are Zeus data, the circles are H1
data and the inverted tri angles the low energy data.}
\label{fig:2sigrho}
\end{figure}

\begin{figure}[htbp] 
\begin{center}
\mbox{\hspace{0.2in}\epsfig{file=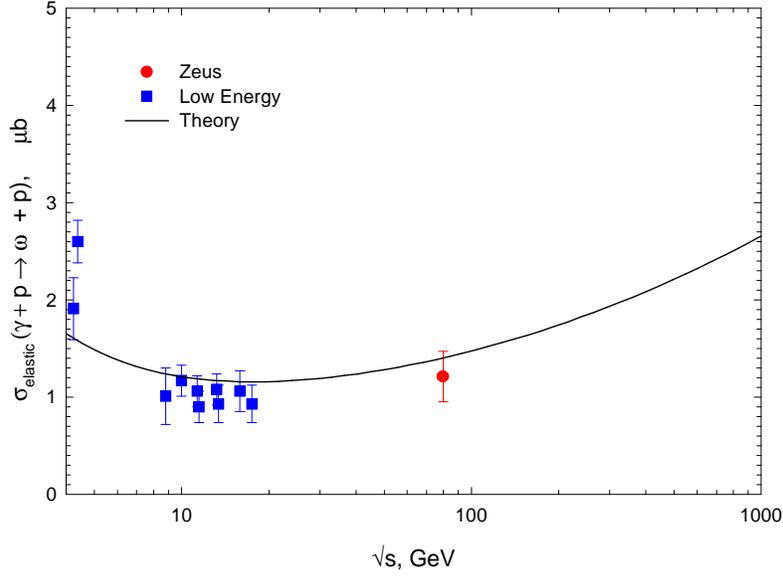,width=4.5in,%
bbllx=90pt,bblly=250pt,bburx=537pt,bbury=560pt,clip=}}
\end{center}
\caption[]{\footnotesize The `elastic' photoproduction cross section, 
$\sigma_{\rm elastic}$, in $\mu$b {\em vs.} $\sqrt s$, in GeV, for the
reaction $\gamma +p\rightarrow\omega + p$. The solid curve is the
predicted cross section, the circles are Zeus data and the squares are
the low energy data.  }
\label{fig:2sigomeg}
\end{figure}

Since we normalize the experimental data to the {\em elastic} cross
section found with $\chi^{\gamma p}$, and {\em not} to
$\frac{1}{2}(\sigma^{\pi^+}_{\rm elastic}+\sigma^{\pi^-}_{\rm
elastic})$, we find that must multiply {\em all}
$\frac{f_{V}^2}{4\pi}$ by 1.65.  Hence, our effective coupling s are
\begin{equation}
 \frac{{f_{\rho}^2}_{\rm eff}}{4\pi}=3.6,\quad
\frac{{f_{\omega}^2}_{\rm eff}}{4\pi}=38.9\quad \rm{and}\quad
\frac{{f_{\phi}^2}_{\rm eff}}{4\pi}=30.4. \label{feff}
\end{equation} 
Thus, we define the $P_{\rm had}^{Vp}$ in \eq{dsdt2gp} and
\eq{sigelgp} as
\begin{equation}
 P_{\rm had}^{Vp}=\frac {4\pi\alpha}{{f_{V}^2}_{\rm eff}}.
\end{equation}
In \fig{fig:2sigrho}, we show our prediction for the `elastic'
reaction $\gamma +p\rightarrow\rho^0 + p$, where we plot the elastic
cross section $\sigma_{\rm elastic}^{\rho p}(s)$ against the cms
energy. The solid curve is the predicted cross section, the squares
are Zeus data, the circles are H 1 data and the inverted triangles the
low energy data.
%
%
In \fig{fig:2sigomeg}, we show our prediction for the `elastic`
reaction $\gamma +p\rightarrow\omega + p$. The solid curve is the
predicted cross section, the circles are Zeus data and the squares are
the low energy data.
The agreement of Figs. (\ref{fig:2sigrho}) and (\ref{fig:2sigomeg}) is
very good, lending further support to the model.

\begin{figure}[htp] 
\begin{center}
\mbox{\epsfig{file=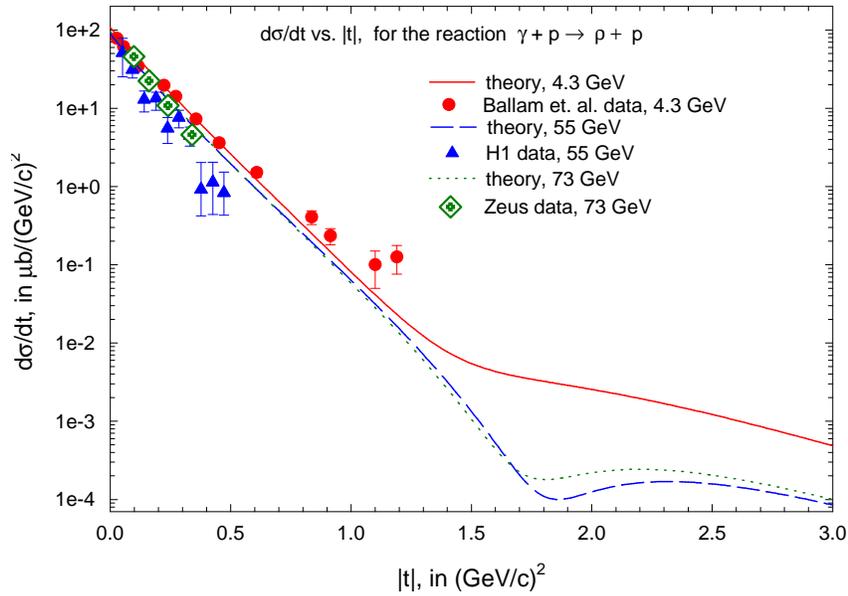,width=4.5in,%
bbllx=100pt,bblly=240pt,bburx=530pt,bbury=550pt,clip=}}
\end{center}
\caption[]{\footnotesize The predicted differential scattering cross section
 $\frac{d\sigma}{dt}$, in $\mu$b/(GeV/c)$^2$ {\em vs.} $|t|$, in
 (GeV/c)$^2$, for the `elastic' reaction $\gamma +p\rightarrow \rho^0+
 p$. The solid curve and the circles (Ballam {\em et al.} data) are at
 $\sqrt s$= 4.3 GeV, the dashed curve and triangles (H1 data) are at
 $\sqrt s$= 55 GeV and the dotted curve and diamonds are at $\sqrt s$=
 73 GeV (Zeus data).}
\label{fig:2dsrho}
\end{figure}

\begin{figure}[htp]
\begin{center}
\mbox{\epsfig{file=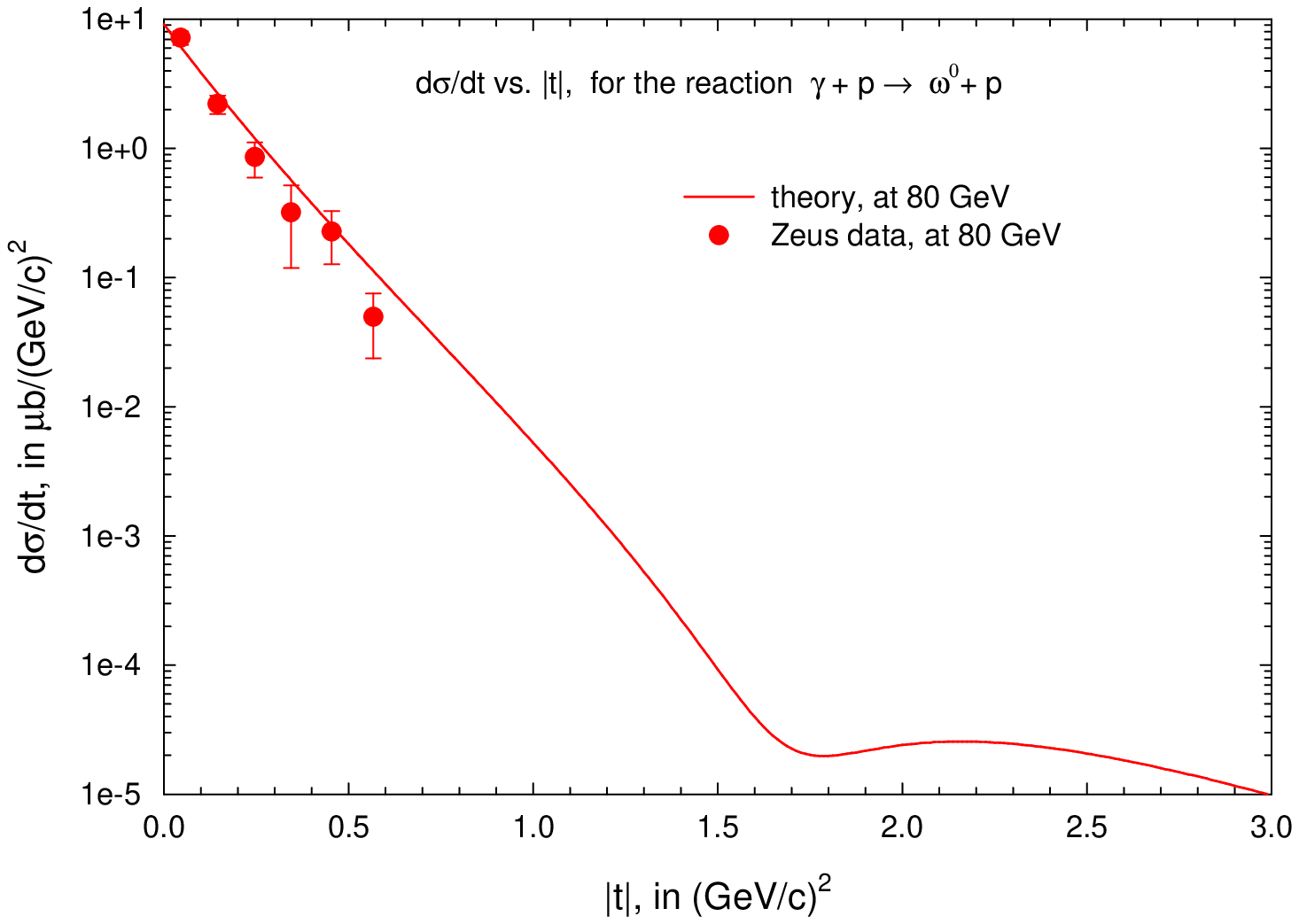,width=4.5in,%
bbllx=80pt,bblly=240pt,bburx=510pt,bbury=550pt,clip=}}
\end{center}
\caption[]{\footnotesize The predicted differential scattering cross section
 $\frac{d\sigma}{dt}$, in $\mu$b/(GeV/c)$^2$ {\em vs.} $|t|$, in
 (GeV/c)$^2$, for the `elastic' reaction $\gamma +p\rightarrow \omega
 + p$, at $\sqrt s$=80 GeV.  The circles are the Zeus data. }
\label{fig:2dsomega}
\end{figure}

\begin{figure}[htbp]
\begin{center}
\mbox{\epsfig{file=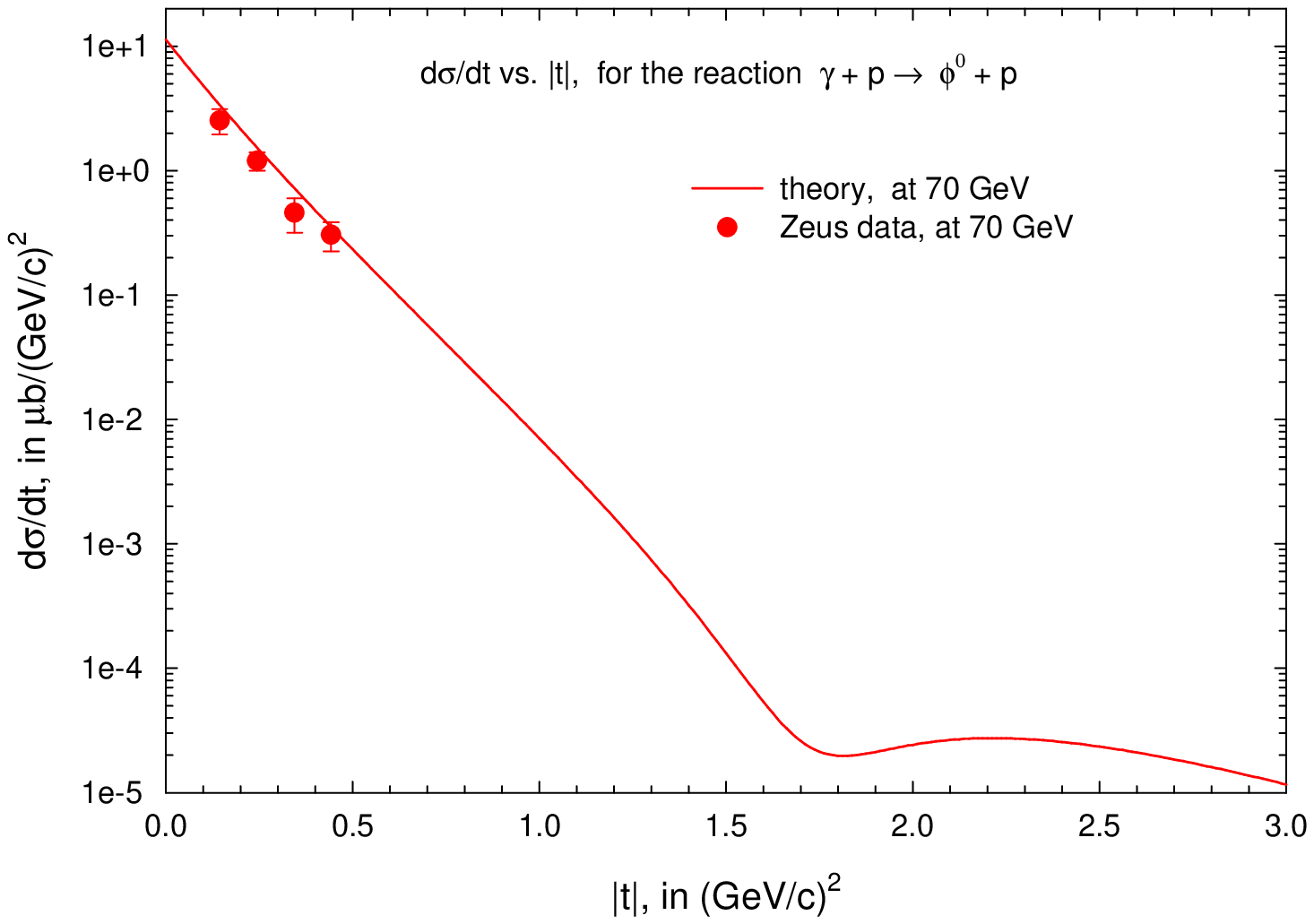,width=4.5in,%
bbllx=40pt,bblly=225pt,bburx=520pt,bbury=570pt,clip=}}
\end{center}
\caption[]{\footnotesize The predicted differential scattering cross section
 $\frac{d\sigma}{dt}$, in $\mu$b/(GeV/c)$^2$ {\em vs.} $|t|$, in
 (GeV/c)$^2$, for the `elastic' reaction $\gamma +p\rightarrow \phi +
 p$, at $\sqrt s$=70 GeV.  The circles are the Zeus data. }
\label{fig:2ds70phi}
\end{figure}

 The predicted differential cross sections, $d\sigma/dt$, for the
 `elastic' reactions $\gamma +p\rightarrow\rho^0 + p$, $\gamma
 +p\rightarrow\omega + p$ and $\gamma +p\rightarrow\phi + p$, for
 diverse energies, are plotted in Figs. \ref{fig:2dsrho},
 \ref{fig:2dsomega} and \ref{fig:2ds70phi}, respec tively.  The
 agreement, in absolute normalization and shape, of the predicted
 differential scattering cross sections with the experimental data for
 all three light mesons for all available energies reinforces even
 more our confidence in our model of $\gamma p$ scattering.
%
%
%
%
%
%
\subsection{How Large is the $\rho$, $\omega$ and $\phi$  Contribution?}
\begin{figure}[htbp] 
\begin{center}
\mbox{\epsfig{file=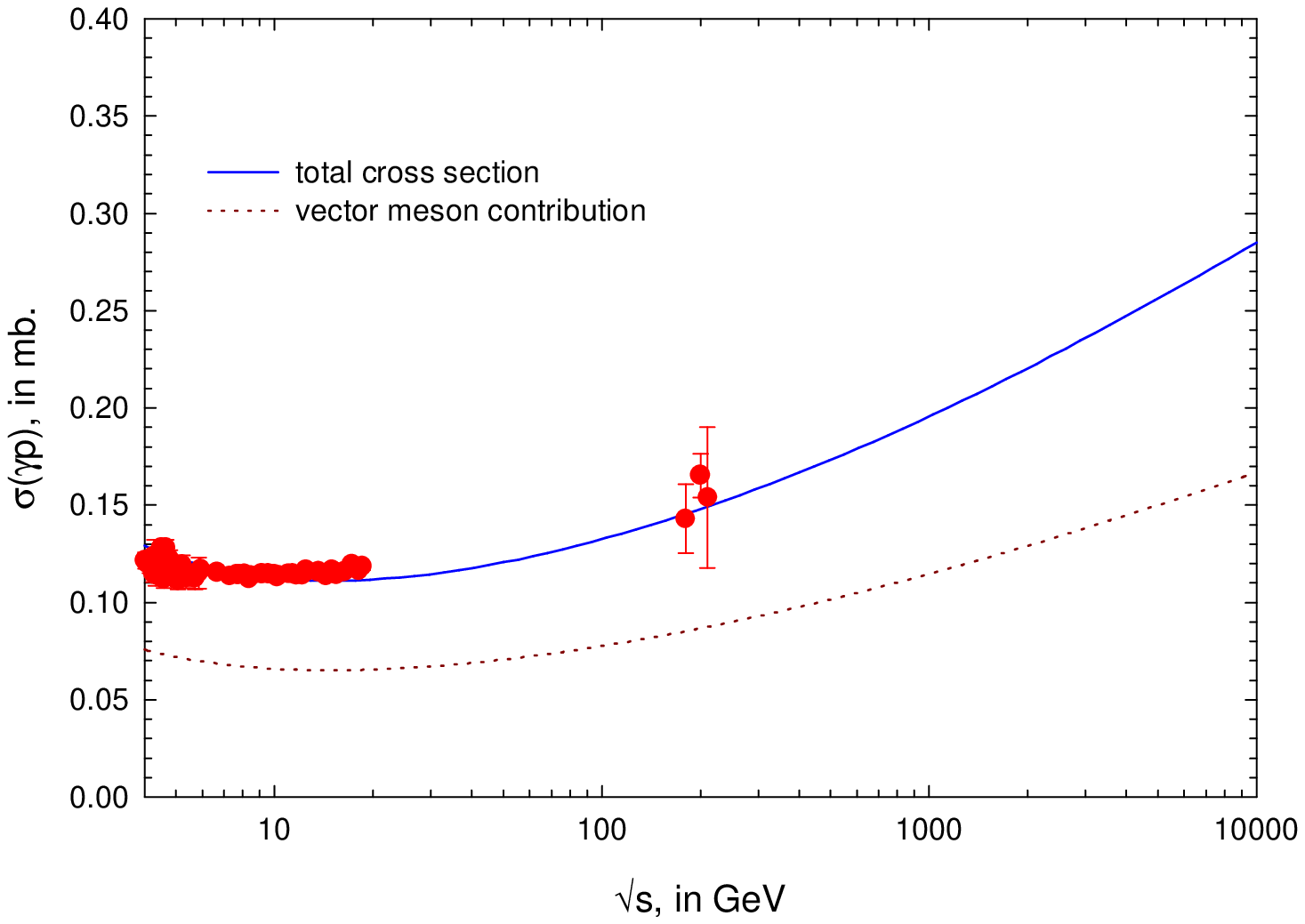,width=4.5in,%
bbllx=100pt,bblly=358pt,bburx=537pt,bbury=660pt,clip=}}
\end{center}
\caption[]{\footnotesize The dotted curve is the prediction for the total
 vector meson photoproduction cross section, in mb, {\em i.e.,} the
 sum of $\gamma +p\rightarrow\rho^0 + p$, $\gamma +p\rightarrow\omega
 + p$ and $\gamma +p\rightarrow\phi + p$, divided by the ratio of
 elastic to total cross se ction (see text). For comparison, we also
 show the solid line and the data which are the predicted and
 experimental total photoproduction cross sections on protons, in mb,
 {\em vs.} $\sqrt s$ in GeV.  }
\label{fig:2sigvec}
\end{figure}

We sum all of our predictions for the elastic vector interactions for
$\rho$, $\omega$ and $\phi$ and divide this sum by the ratio of
$\sigma_{\rm elastic}/\sigma_{\rm tot}$ (obtained from $\chi^{\gamma
p}$, using \eq{sigtot} and \eq{sigel}). We call this quantity the {\rm
total} vector meson con tribution. In \fig{fig:2sigvec}, we compare
this to the total $\gamma p$ cross section.  We find that the fraction
of the total cross section for $\gamma p$ reactions that is
contributed by the three light vector mesons ($\rho$, $\omega$ and
$\phi$) is $\approx 0.60$. The remaining 40\% could, in the spirit of
VMD, be made up of heavier vector meson states, or perhaps could be
continuum states, or, indeed, a mixture of both.

\section{$\gamma\gamma$ Interactions\label{gammagamma}}
In this Section, we consider $\gamma\gamma$ interactions. As we did
for $\gamma p$ interactions, we will take the eikonal $\chi^{\gamma
p}(s,b)$ and again multiply every cross section by 2/3 and multiply
each $\mu$ by $\sqrt {3/2}$. Thus, we have
\begin{eqnarray}
\chi^{\gamma \gamma}(s,b) 
&=& i\left [\frac{4}{9}
\sigma_{qq}(s)W\left(b;\frac{3}{2}\mu_{qq}\right) + \frac{4}{9}
\sigma_{qg}(s)W\left(b;\frac{3}{2}\sqrt{\mu_{qq}\mu_{gg}}\right)\right.\nonumber\\
&&\qquad \left. + \frac{4}{9}
\sigma_{gg}(s)W\left(b;\frac{3}{2}\mu_{gg}\right)\right ]\,
.\label{chigammagamma}
\end{eqnarray}
\subsection{$\gamma\gamma$ Total Cross Section Prediction\label{sigtotgg}}
\begin{figure}[htbp]
\begin{center}
\mbox{\epsfig{file=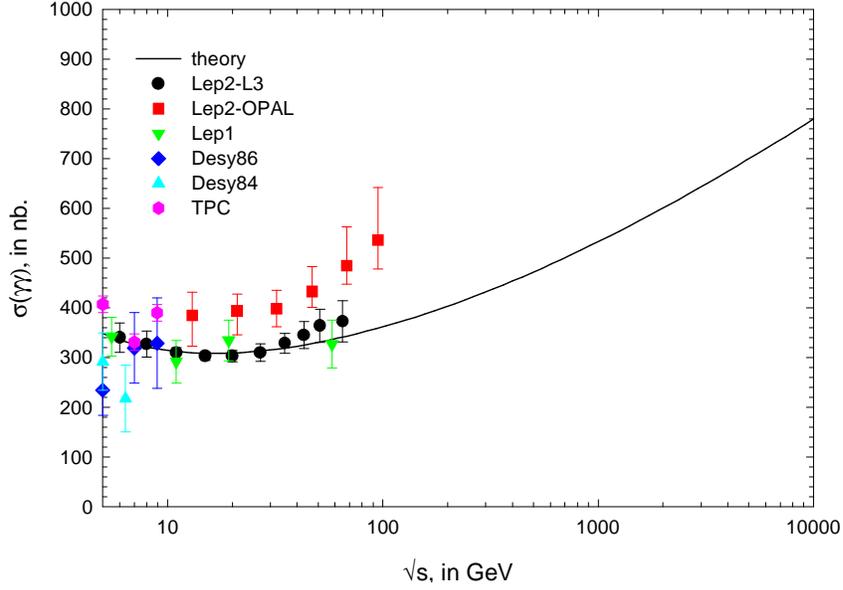,width=4.5in,%
bbllx=100pt,bblly=358pt,bburx=537pt,bbury=660pt,clip=}}
\end{center}
\caption[]{\footnotesize The predicted total cross section, $\sigma_{\rm tot}$, 
in nb {\em vs.} $\sqrt s$, in GeV,  for $\gamma \gamma$ scattering. 
The data sources are indicated in the legend.}
\label{fig:3sigtot}
\end{figure}

Again, using vector dominance and the $\gamma \gamma$ eikonal of
\eq{chigammagamma}, we can now write, using \eq{sigtot},
\begin{eqnarray}
\sigma_{\rm tot}^{\gamma \gamma}(s)
=&2\left((P_{\rm had}\right)^2\int\,\left[1-e^{-{\chii}^{\gamma
\gamma} (b,s)}\cos({\chir}^{\gamma p}
(b,s))\right]\,d^2\vec{b},\label{sigtotgammagamma}
\end{eqnarray}
where, again, $P_{\rm had}=1/240$ is the probability that a photon
will interact as a hadron.  In \fig{fig:3sigtot} we plot our
prediction for $\sigma_{\rm tot}^{\gamma \gamma}(s)$ as a function of
the cms energy and compare it to the various sets of experimental
data. It is clear that VMD selects the L3 and casts doubt on the
preliminary OPAL results\cite{exp}.
\subsection{`Elastic' $\gamma\gamma$ Reactions\label{sec:elasticgg}}
In this Section, we make prediction for `elastic' $\gamma\gamma$
reactions, in which {\em both} photons turn into vector mesons, and
then elastically scatter off each other, {\em i.e.,}
$\gamma+\gamma\rightarrow V_i+V_j\rightarrow V_i+V_j$, where
$V_i,V_j=\rho,\,\omega,\,\phi$.  We consider here the 6 reactions
\begin{eqnarray}
\gamma + \gamma \rightarrow \rho^0 +\rho^0,\label{rhorho}\\
\gamma + \gamma \rightarrow \rho^0 +\omega,\label{rhoomega}\\
\gamma + \gamma \rightarrow \rho^0 +\phi,\label{rhophi}\\
\gamma + \gamma \rightarrow \omega +\omega,\label{omegaomega}\\
\gamma + \gamma \rightarrow \omega +\phi,\label{omegaphi}\\
\gamma + \gamma \rightarrow \phi +\phi.\label{phiphi}
\end{eqnarray} 
There currently exist no data for such reactions, but hopefully, there
will be in the foreseeable future---perhaps these predictions will be
useful for experimental planning.
\subsubsection{`Elastic' $\gamma \gamma$ Cross Sections\label{sec:sigelgg}}
To find the total `elastic' scattering cross sections, we invoke
\eq{sigel} of Appendix \ref{app:sigel}, using the eikonal
$\chi^{\gamma\gamma}$ of \eq{chigammagamma}, multiplied by the factors
$P_{had}^{V_i}P_{had}^{V_j}$, {\em i.e.,} $\sigma_{\rm el}(s)$ as
\begin{eqnarray}
\sigma_{\rm elastic}^{\gamma\gamma}(s)
&=&2P_{had}^{V_i}P_{had}^{V_j}\int\left|1-e^{i\chi^{\gamma\gamma}(b,s)}
\right|^2\,d^2\vec{b},
\qquad {\rm if\ } i\neq j,\label{different}\\
&=&\phantom{2}P_{had}^{V_i}P_{had}^{V_j}\int\left|1-e^{i\chi^{\gamma\gamma}(b,s)}
\right|^2\,d^2\vec{b},
\qquad {\rm if\ } i=j.\label{same}
\label{sigelgg}
\end{eqnarray}
The factor of 2 in \eq{different}, where there are {\em unlike} mesons
in the final state, takes into account, for example, that {\em either}
photon in \eq{rhoomega} could turn into a $\rho^0$. In
eqns. (\ref{different}) and (\ref{same}), the factor $P_{\rm
had}^{V_i}=\frac{4\pi\alpha}{{f_{V_i}^2} _{\rm eff}}, \ V_i=\rho,\
\omega,\ \phi$, where, from \eq{feff}, $ \frac{{f_{\rho}^2}_{\rm
eff}}{4\pi}=3.6,\
\frac{ {f_{\omega}^2}_{\rm eff}}{4\pi}=38.9,$ 
and $\frac{{f_{\phi}^2}_{\rm eff}}{4\pi}=30.4$.

\begin{figure}[htp]%
\begin{center}
\mbox{\epsfig{file=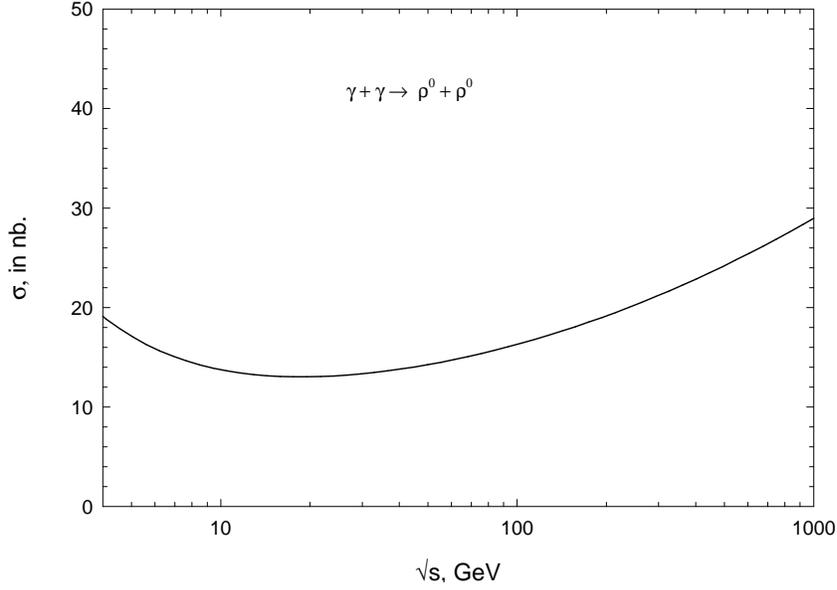,width=4.5in,%
bbllx=100pt,bblly=358pt,bburx=537pt,bbury=660pt,clip=}}
\end{center}
\caption[]{\footnotesize The predicted  cross section in nb {\em vs.} 
$\sqrt s$, in GeV, for the `elastic' reaction $\gamma
+\gamma\rightarrow \rho^0 + \rho^0$. }
\label{fig:3sigrr}
\end{figure}

\begin{figure}[htbp]%
\begin{center}
\mbox{\epsfig{file=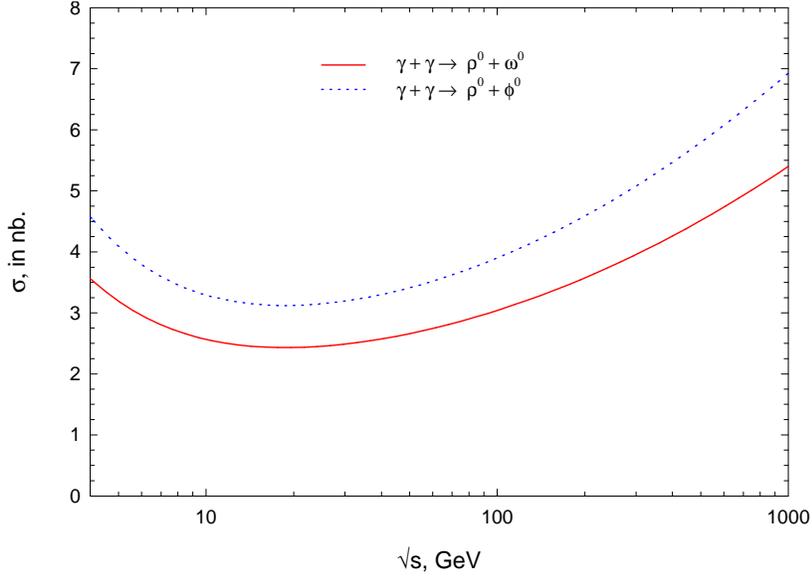,width=4.5in,%
bbllx=85pt,bblly=250pt,bburx=530pt,bbury=550pt,clip=}}
\end{center}
\caption[]{\footnotesize The predicted  cross sections in nb {\em vs.} 
$\sqrt s$, in GeV, for $\gamma\gamma$ `elastic' reactions. The solid
curve is for the reaction $\gamma +\gamma\rightarrow \rho^0 + \omega$
and the dotted curve for $\gamma +\gamma\rightarrow \rho^0 + \phi$.}
\label{fig:3sigro}
\end{figure}

\begin{figure}[htbp]%
\begin{center}
\mbox{\epsfig{file=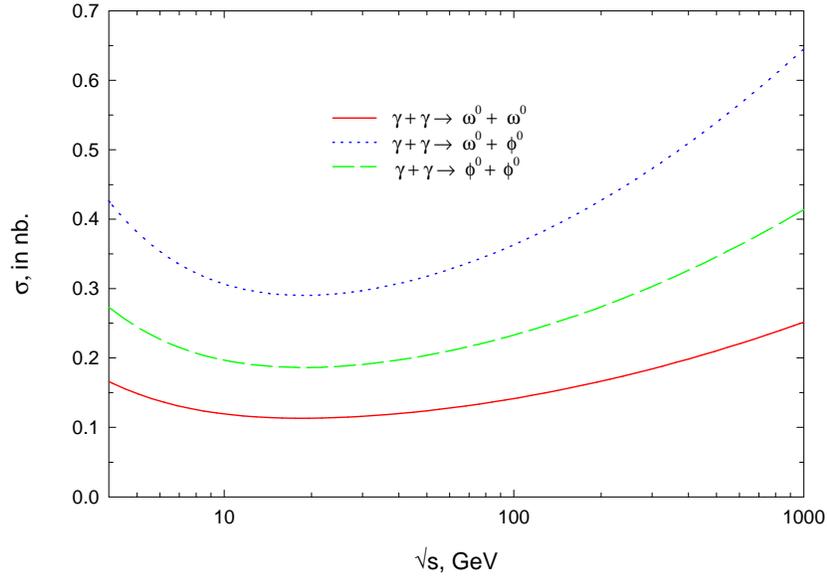,width=4.5in,%
bbllx=90pt,bblly=260pt,bburx=537pt,bbury=560pt,clip=}}
\end{center}
\caption[]{\footnotesize The predicted  cross sections in nb {\em vs.} 
$\sqrt s$, in GeV, for $\gamma\gamma$ `elastic' reactions. The solid
curve is for the reaction $\gamma +\gamma\rightarrow \omega + \omega$,
the dotted curve for $\gamma +\gamma\rightarrow \omega + \phi$ and the
dashed curve is f or $\gamma +\gamma\rightarrow \phi + \phi$.}
\label{fig:3sigoo}
\end{figure}

We show in \fig{fig:3sigrr} the predicted cross section for reaction
(\ref{rhorho}), in \fig{fig:3sigro} for reactions (\ref{rhoomega}) and
(\ref{rhophi}) and finally, in \fig{fig:3sigoo}, for reactions
(\ref{omegaomega}), (\ref{omegaphi}) and (\ref{phiphi}), as a function
of $\sqrt s$.
\newpage

\subsubsection{Slope Parameters $B$}
\begin{figure} [htbp]
\begin{center}
\mbox{\epsfig{file=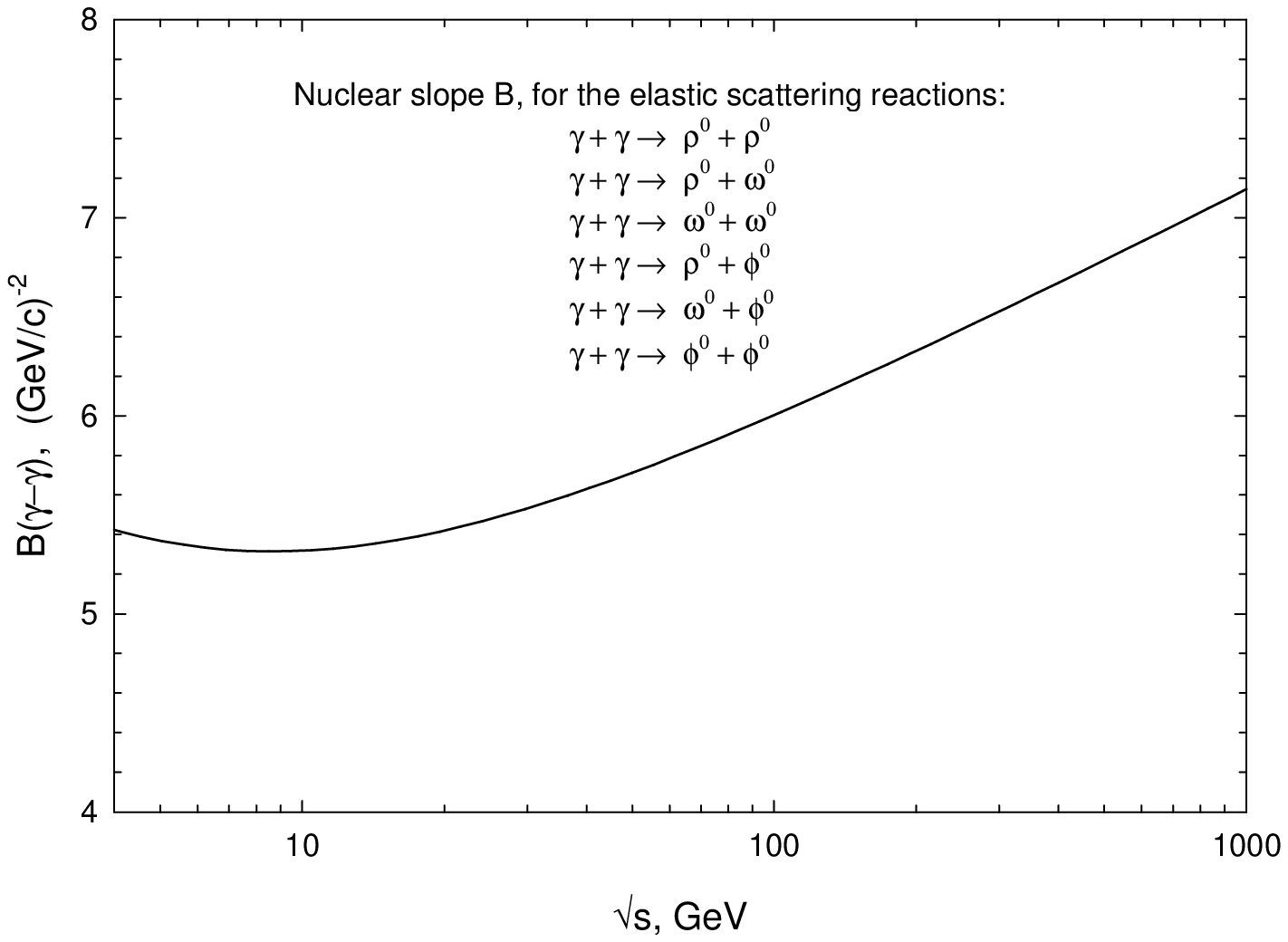,width=4.5in,%
bbllx=90pt,bblly=250pt,bburx=537pt,bbury=560pt,clip=}}
\end{center}
\caption[]{\footnotesize The predicted nuclear slope parameter B, 
in (GeV/c)$^{-2}$ {\em vs.} $\sqrt s$, in GeV, for the `elastic '
$\gamma\gamma$ reactions $\gamma +\gamma\rightarrow \rho^0 + \rho^0$,
$\gamma +\gamma\rightarrow \rho^0 + \omega$, $\gamma
+\gamma\rightarrow \omega^0 + \omega^0$, $\gamma +\gamma\rightarrow
\rho^0 + \phi$, $\gamma +\gamma \rightarrow
\omega + \phi$ and$\gamma +\gamma \rightarrow \phi + \phi$. }
\label{fig:3bvec}
\end{figure}

Using \eq{Bfinal} and the eikonal $\chi^{\gamma\gamma}$, we predict
the nuclear slope parameter for the `elastic' reaction of
eqns. (\ref{rhorho})--(\ref{phiphi}), as a function of energy.  Of
course, the slopes are the same for {\em all} elastic $\gamma\gamma$
reactions. The predicted slopes B, a s a function of $\sqrt s$, are
shown in \fig{fig:3bvec}.
%
\subsubsection{Differential Cross Sections\label{sec:dsdtgg}}
Using \eq{dsdt2} and $\chi^{\gamma\gamma}$ of \eq{chigammagamma}, we
can write
\begin{eqnarray}
\frac{d\sigma}{dt}(s,t)&=&2P_{had}^{V_i}P_{had}^{V_j}\frac{1}{4\pi}\left|
\int J_0(qb)(1-e^{i\chi^{\gamma\gamma}(b,s)})\,d^2\vec{b}\,\right|^2,
\qquad {\rm if\ } i\neq j,\label{differentdsdt}\\
&=&\phantom{2}P_{had}^{V_i}P_{had}^{V_j}\frac{1}{4\pi}\left|\int
J_0(qb)(1-e^{i\chi^{\gamma\gamma}(b,s)})\,d^2\vec{b}\,\right|^2,
\qquad {\rm if\ } i=j,\label{samedsdt}
\label{dsdt2gg}
\end{eqnarray}
where $t=-q^2$.  The factor of 2 in \eq{differentdsdt}, where there
are {\em unlike} mesons in the final state, again takes into account,
for example, that {\em either} photon in \eq{rhoomega} could turn into
a $\rho^0$. In eqns. (\ref{differentdsdt}) and (\ref{samedsdt}), the
factor $P_{\rm had}^{V_i}=\frac{4\pi\ alpha}{{f_{V_i}^2}_{\rm eff}}, \
V_i=\rho,\ \omega,\ \phi$, where, from \eq{feff}, $
\frac{{f_{\rho}^2}_{\rm eff}}{4\pi}=3.6,\
\frac{ {f_{\omega}^2}_{\rm eff}}{4\pi}=38.9,$ 
and $\frac{{f_{\phi}^2}_{\rm eff}}{4\pi}=30.4$.

\begin{figure}[htbp]
\begin{center}
\mbox{\epsfig{file=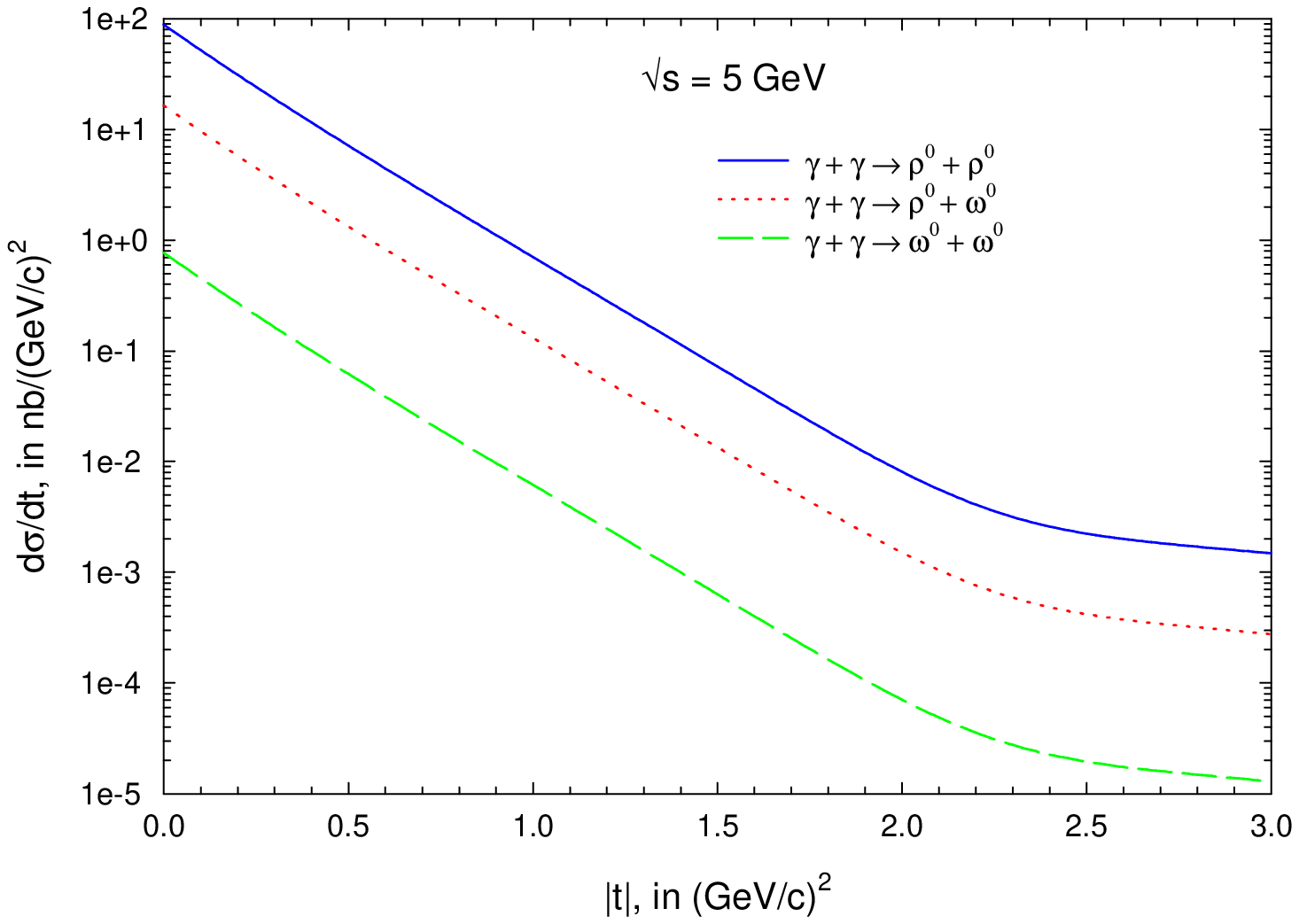,width=4.5in,%
bbllx=75pt,bblly=245pt,bburx=537pt,bbury=560pt,clip=}}
\end{center}
\caption[]{\footnotesize The predicted differential scattering 
cross section $\frac{d\sigma}{dt}$, in nb/(GeV/c)$^2$ {\em vs.} $|t|$,
in (GeV/c)$^2$, for the `elastic' $\gamma\gamma$ reactions, at $\sqrt
s$=5 GeV. The solid curve is for the reaction $\gamma
+\gamma\rightarrow \rho^0 + \rho^0$, the d otted curve for $\gamma
+\gamma\rightarrow \rho + \omega$ and the dashed curve for $\gamma
+\gamma\rightarrow \omega + \omega$.}
\label{fig:3ds5rr}
\end{figure}

In \fig{fig:3ds5rr} we show the predicted differential scattering
cross section $\frac{d\sigma}{dt}$ as a function of $|t|$ for the
`elastic' $\gamma\gamma$ reactions of eqns. (\ref{rhorho}),
(\ref{rhoomega}), and (\ref{omegaomega}), at $\sqrt s$=5 GeV. The
solid curve is for the reaction $\gamma +\gamma\rightarrow \rho^0 +
\rho^0$, the dotted curve for $\gamma +\gamma\rightarrow \rho^0 +
\omega$ and the dashed curve for $\gamma +\gamma\rightarrow \omega +
\omega$.

\begin{figure}[htbp]
\begin{center}
\mbox{\epsfig{file=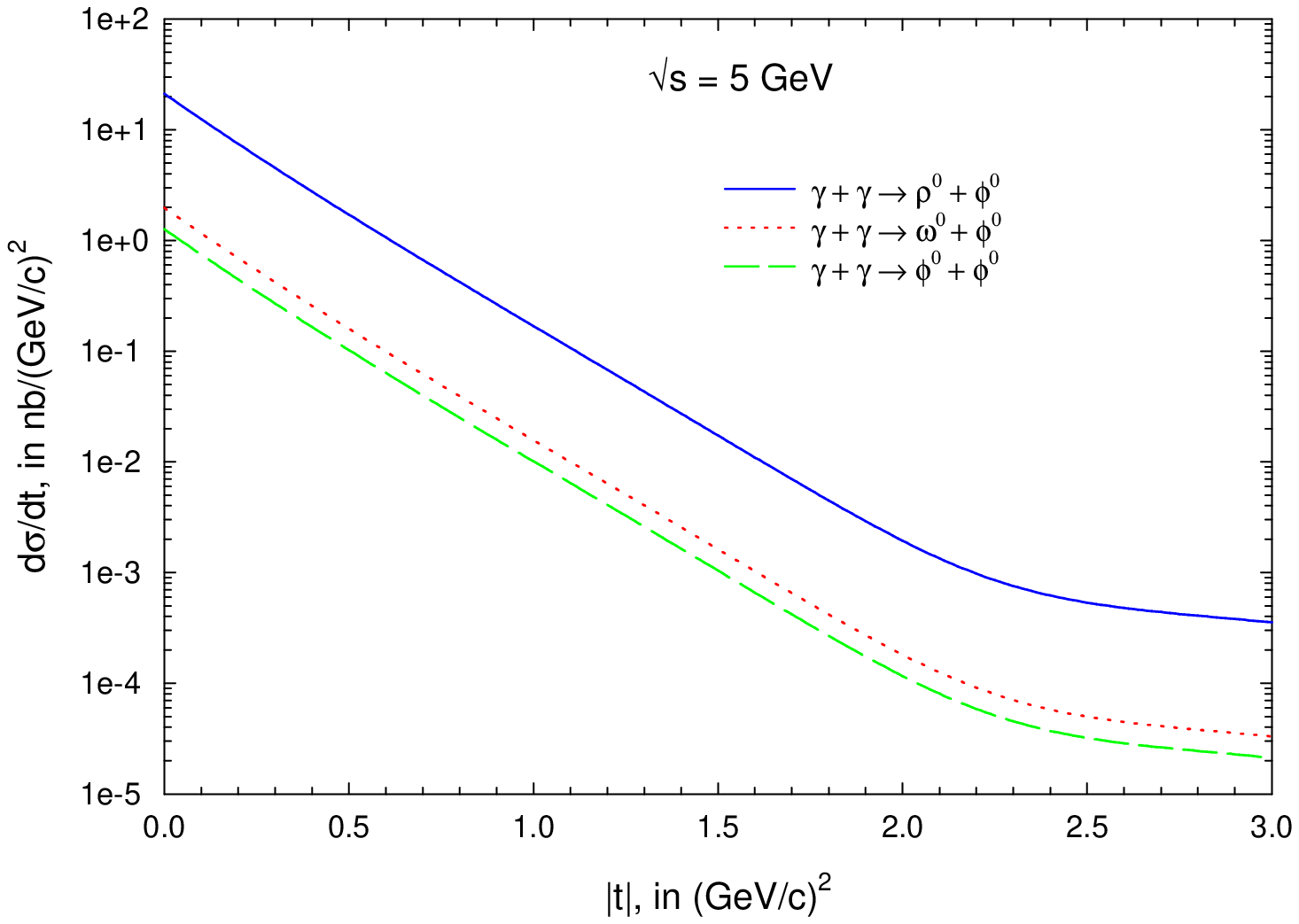,width=4.5in,%
bbllx=80pt,bblly=250pt,bburx=537pt,bbury=560pt,clip=}}
\end{center}
\caption[]{\footnotesize The predicted differential scattering cross 
section $\frac{d\sigma}{dt}$, in nb/(GeV/c)$^2$ {\em vs.} $|t|$, in
(GeV/c)$^2$, for the `elastic' $\gamma\gamma$ reactions, at $\sqrt
s$=5 GeV. The solid curve is for the reaction $\gamma
+\gamma\rightarrow \rho^0 + \phi$, the dot ted curve for $\gamma
+\gamma\rightarrow \omega + \phi$ and the dashed curve for $\gamma
+\gamma\rightarrow \phi + \phi$.}
\label{fig:3ds5rp}
\end{figure}

In \fig{fig:3ds5rp} we show the predicted differential scattering
cross section $\frac{d\sigma}{dt}$ for the `elastic' $\gamma\gamma$
reactions of eqns. (\ref{rhophi}), (\ref{omegaphi}), and
(\ref{phiphi}), at $\sqrt s$=5 GeV. The solid curve is for the
reaction $\gamma +\gamma\rightarrow \rho^0 + \rho^0$, the dotted curve
for $\gamma +\gamma\rightarrow \rho^0 + \omega$ and the dashed curve
for $\gamma +\gamma\rightarrow \omega + \omega$.
%

\begin{figure}[htbp]
\begin{center}
\mbox{\epsfig{file=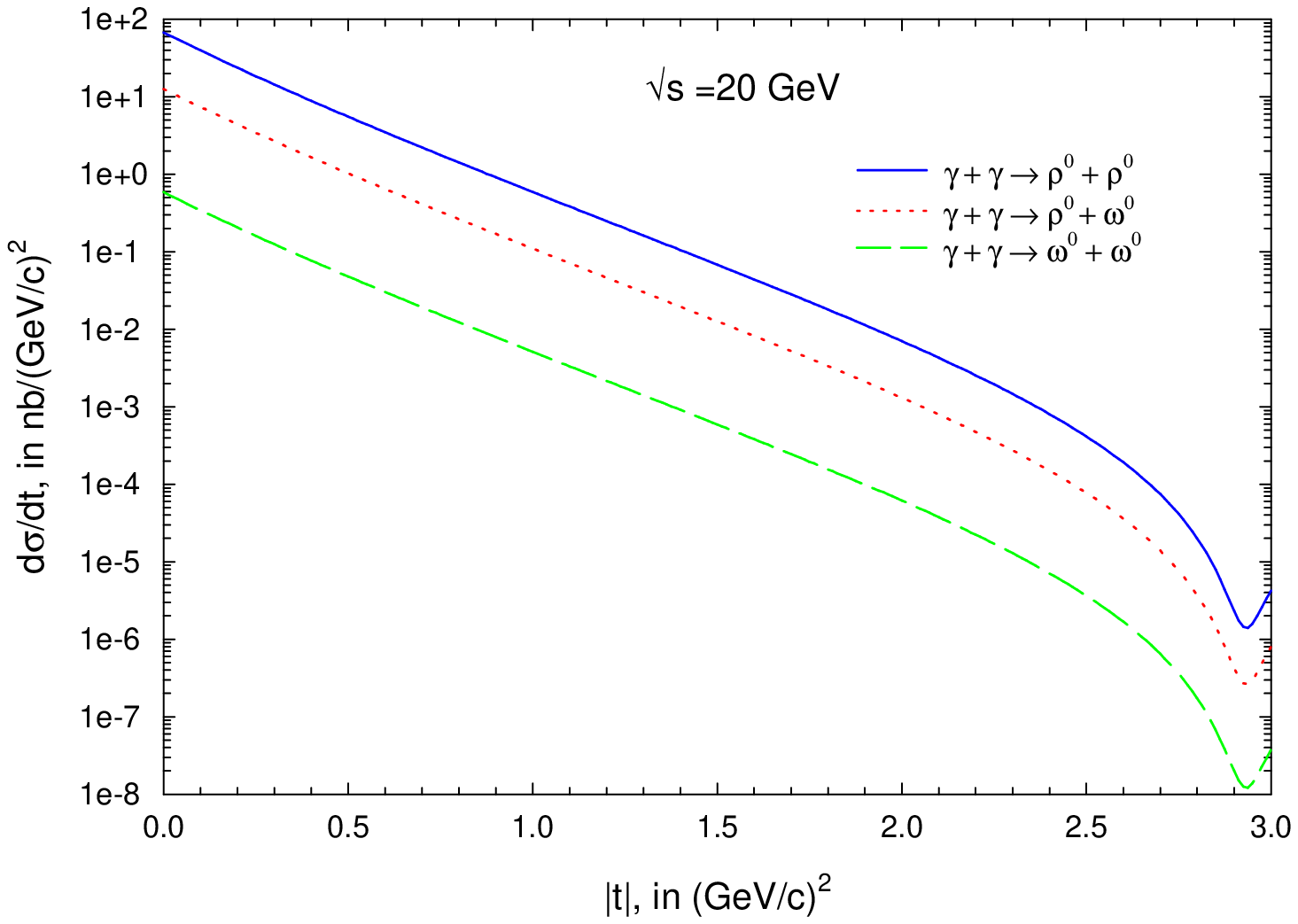,width=4.5in,%
bbllx=80pt,bblly=250pt,bburx=537pt,bbury=560pt,clip=}}
\end{center}
\caption[]{\footnotesize The predicted differential scattering 
cross section $\frac{d\sigma}{dt}$, in nb/(GeV/c)$^2$ {\em vs.} $|t|$,
in (GeV/c)$^2$, for the `elastic' $\gamma\gamma$ reactions, at $\sqrt
s$=20 GeV. The solid curve is for the reaction $\gamma
+\gamma\rightarrow \rho^0 + \rho^0$, the dotted curve for $\gamma
+\gamma\rightarrow \rho^0 + \omega$ and the dashed curve for $\gamma
+\gamma\rightarrow \omega + \omega$.}
\label{fig:3ds20rr}
\end{figure}

In \fig{fig:3ds20rr} we show the predicted differential scattering
cross section $\frac{d\sigma}{dt}$ for the `elastic' $\gamma\gamma$
reactions of eqns. (\ref{rhorho}), (\ref{rhoomega}), and
(\ref{omegaomega}), at $\sqrt s$=20 GeV. The solid curve is for the
reaction $\gamma +\gamma\rightarrow \rho^0 + \rho^0$, the dotted
curve for $\gamma +\gamma\rightarrow \rho^0 + \omega$ and the dashed
curve for $\gamma +\gamma\rightarrow \omega + \omega$.

\begin{figure}[htbp]
\begin{center}
\mbox{\epsfig{file=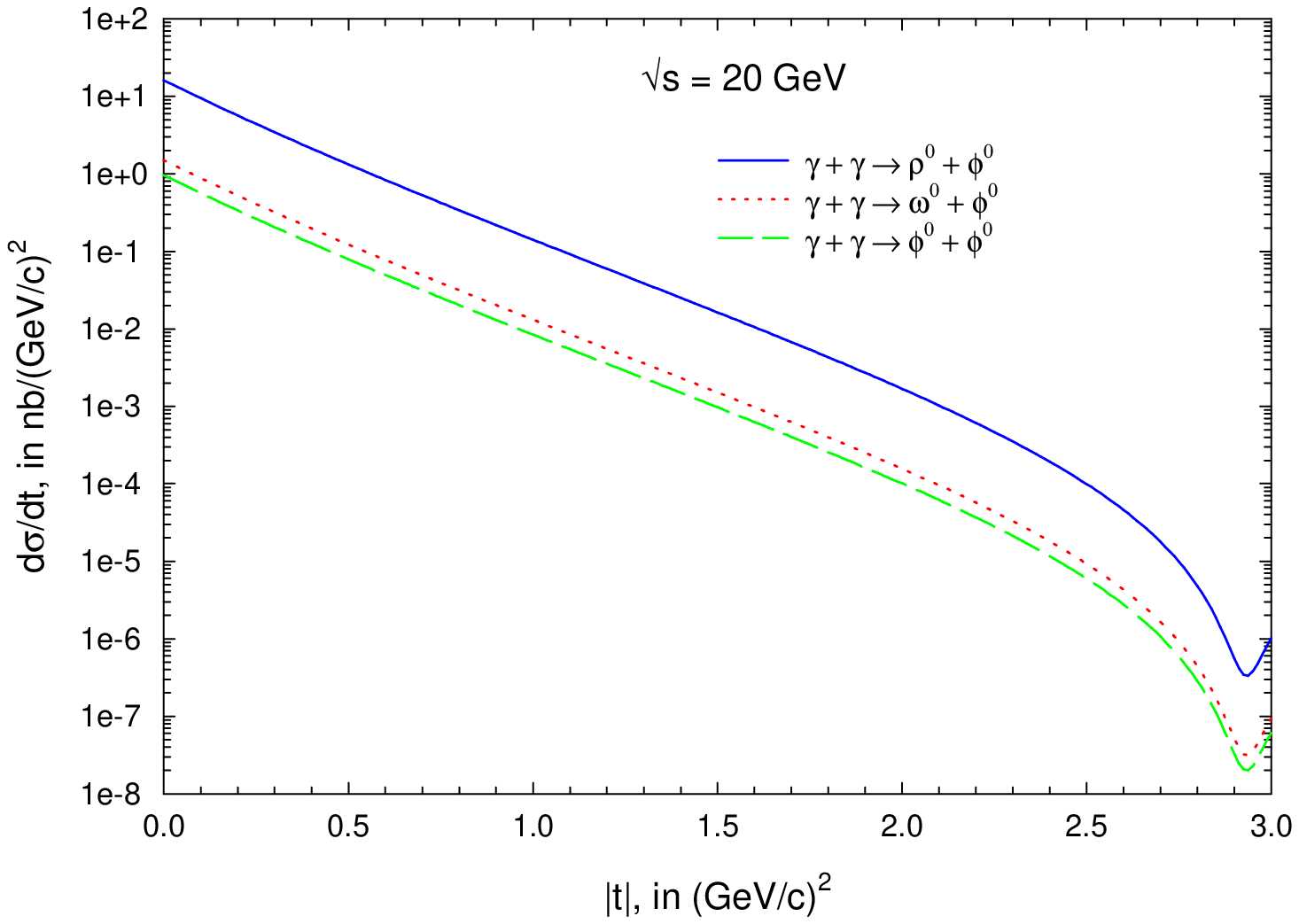,width=4.5in,%
bbllx=80pt,bblly=250pt,bburx=537pt,bbury=560pt,clip=}}
\end{center}
\caption[]{\footnotesize The predicted differential scattering cross section 
$\frac{d\sigma}{dt}$, in nb/(GeV/c)$^2$ {\em vs.} $|t|$, in
(GeV/c)$^2$, for the `elastic' $\gamma\gamma$ reactions, at $\sqrt
s$=20 GeV. The solid curve is for the reaction $\gamma
+\gamma\rightarrow \rho^0 + \phi^0$, the dotted curve for $\gamma
+\gamma\rightarrow \omega^0 + \phi^0$ and the dashed curve for $\gamma
+\gamma\rightarrow \phi^0 + \phi^0$.}
\label{fig:3ds20rp}
\end{figure}

In \fig{fig:3ds20rp} we show the predicted differential scattering
cross section $\frac{d\sigma}{dt}$ for the `elastic' $\gamma\gamma$
reactions of eqns. (\ref{rhophi}), (\ref{omegaphi}), and
(\ref{phiphi}), at $\sqrt s$=20 GeV. The solid curve is for the
reaction $\gamma +\gamma\rightarrow \rho^ 0 + \rho^0$, the dotted
curve for $\gamma +\gamma\rightarrow \rho^0 + \omega$ and the dashed
curve for $\gamma +\gamma\rightarrow \omega + \omega$.
%

\begin{figure}[htbp]
\begin{center}
\mbox{\epsfig{file=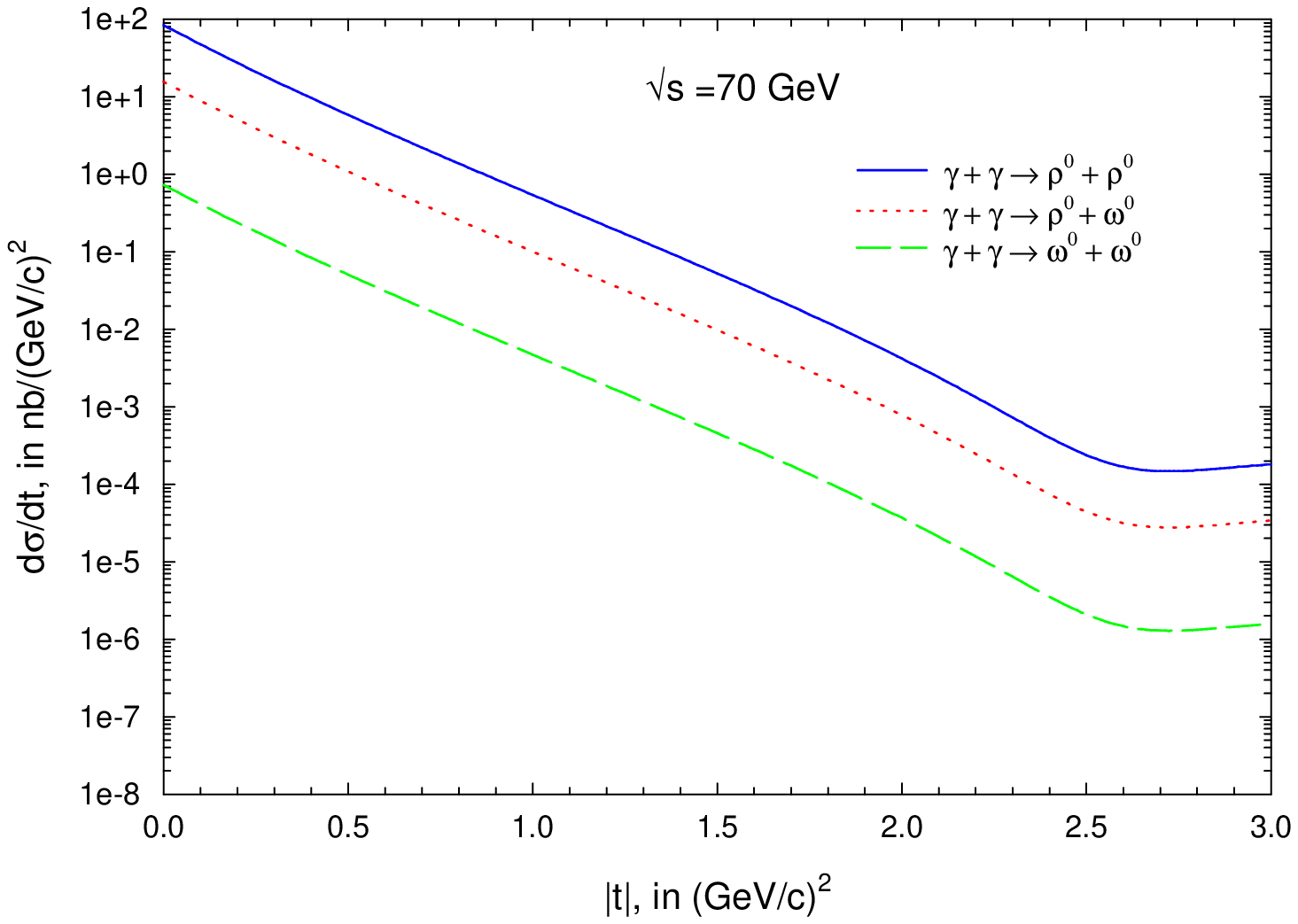,width=4.5in,%
bbllx=90pt,bblly=250pt,bburx=537pt,bbury=560pt,clip=}}
\end{center}
\caption[]{\footnotesize The predicted differential scattering cross section
 $\frac{d\sigma}{dt}$, in nb/(GeV/c)$^2$ {\em vs.} $|t|$, in
 (GeV/c)$^2$, for the `elastic' $\gamma\gamma$ reactions, at $\sqrt
 s$=70 GeV. The solid curve is for the reaction $\gamma
 +\gamma\rightarrow \rho^0 + \rho^0$, the dotted curve for $\gamma
 +\gamma\rightarrow \rho^0 + \omega^0$ and the dashed curve for
 $\gamma +\gamma\rightarrow \omega^0 + \omega^0$.}
\label{fig:3ds70rr}
\end{figure}

In \fig{fig:3ds70rr} we show the predicted differential scattering
cross section $\frac{d\sigma}{dt}$ for the `elastic' $\gamma\gamma$
reactions of eqns. (\ref{rhorho}), (\ref{rhoomega}), and
(\ref{omegaomega}), at $\sqrt s$=70 GeV. The solid curve is for the
reaction $\gamma +\gamma\rightarrow \rho^0 + \rho^0$, the dotted
curve for $\gamma +\gamma\rightarrow \rho^0 + \omega^0$ and the dashed
curve for $\gamma +\gamma\rightarrow \omega^0 + \omega^0$.

\begin{figure}[htbp]
\begin{center}
\mbox{\epsfig{file=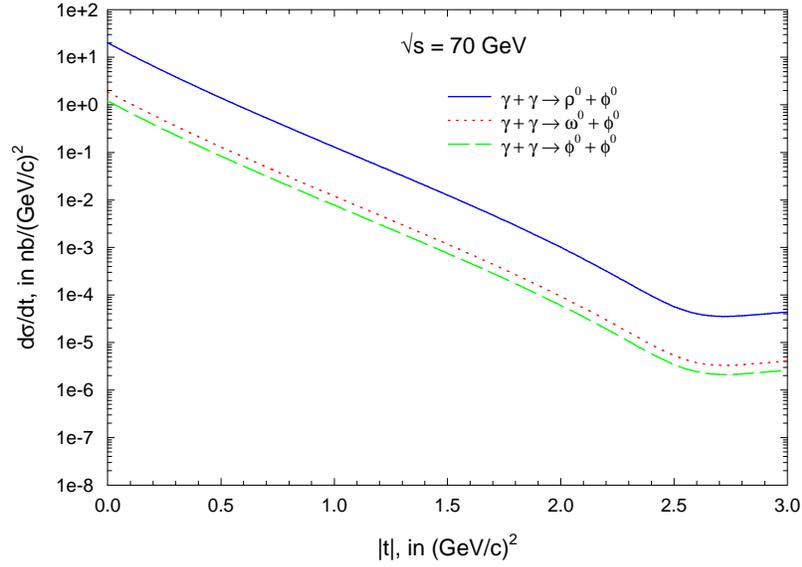,width=4.5in,%
bbllx=80pt,bblly=250pt,bburx=537pt,bbury=560pt,clip=}}
\end{center}
\caption[]{\footnotesize The predicted differential scattering cross section $\frac{d\sigma}{dt}$, in nb/(GeV/c)$^2$ {\em vs.} $|t|$, in (GeV/c)$^2$, for the `elastic' $\gamma\gamma$ reactions, at $\sqrt s$=70 GeV. The solid curve is for the reaction $\gamma +\gamma\rightarrow \rho^0 + \phi$, the dotted curve for $\gamma +\gamma\rightarrow \omega + \phi$ and the
dashed curve for $\gamma +\gamma\rightarrow \phi + \phi$.}
\label{fig:3ds70rp}
\end{figure}

In \fig{fig:3ds70rp} we show the predicted differential scattering
cross section $\frac{d\sigma}{dt}$ for the `elastic' $\gamma\gamma$
reactions of eqns. (\ref{rhophi}), (\ref{omegaphi}), and
(\ref{phiphi}), at $\sqrt s$=20 GeV. The solid curve is for the
reaction $\gamma +\gamma\rightarrow \rho^ 0 + \rho^0$, the dotted
curve for $\gamma +\gamma\rightarrow \rho^0 + \omega$ and the dashed
curve for $\gamma +\gamma\rightarrow \omega + \omega$.

It would be most interesting to be able to measure the predicted $|t|$
structure shown in these differential cross section curves.

\section{Summary and Conclusions\label{sec:conclusions}}
\begin{figure}[htbp]
\begin{center}
\mbox{\epsfig{file=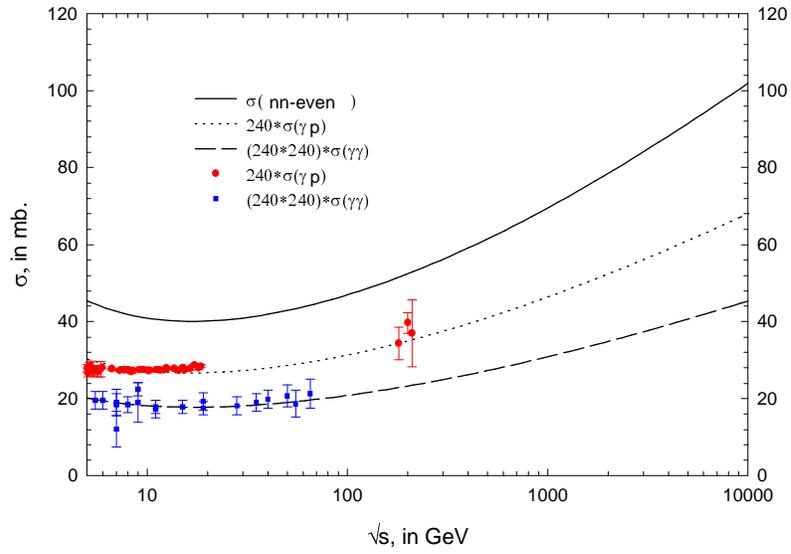,width=4.5in,%
bbllx=80pt,bblly=300pt,bburx=550pt,bbury=630pt,clip=}}
\end{center}
\caption[]{\footnotesize The solid curve is the fitted total cross section 
for nucleon-nucleon scattering, using the even eikonal . The dotted
curve is the predicted total cross section for $\gamma p$ scattering
multiplied by $1/P_{\rm had}$ (=240). The circles are the $\gamma p$
data multiplied by $1/P_{\rm had}$. The dashed curve is the predicted
total cross section for $\gamma \gamma$ scattering multiplied by
$(1/P_{\rm had})^2$ (=240*240). The squares are $\gamma \gamma$ total
cross section data multiplied by $(1/P_{\rm had})^2$.}
\label{fig:123sig}
\end{figure}

Our conclusions for the total cross section for $\gamma p$ and $\gamma
\gamma$ are summarized in \fig{fig:123sig}. In order to scale
nucleon-nucleon, $\gamma p$ and $\gamma\gamma$ cross sections to a
common curve, we have multiplied the $\gamma p$ cross sections by
$1/P_{\rm had}$ (=240) and the $
\gamma\gamma$ cross sections by $(1/P_{\rm had})^2$ (=240$^2$). 
The nucleon-nucleon calculation is made using the even eikonal.  For
clarity, we have not included the Opal $\gamma\gamma$ experimental
data. Basically, both the data and our theory approximately satisfy
factorization, with
\be
\frac{\sigma_{\rm tot}^{\rm nn-even}}{\sigma_{\rm tot}^{\gamma p}} =
\frac{\sigma_{\rm tot}^{\gamma p}}{\sigma_{\rm tot}^{\gamma \gamma}},\label{fact}
\ee
an immediate consequence of the eikonal being small in the energy
region considered (up to $\approx 2$ TeV). The small eikonal we find
is consistent with the Tevatron energy not yet being in `asymptopia'.

All data are in agreement with our QCD-inspired eikonal model, which
requires that we use the even eikonal. When appropriate factors of 2/3
(for quark counting) are introduced, and an energy independent factor
$P_{\rm had}=1/240$ is introduced, we have a natural explanation of
$\gamma p$ interacti ons.  Finally, when we again introduce another
factor of 2/3, and the factor $P_{\rm had}^2 =\frac{1}{(240^2)}$, we
can explain the total $\gamma\gamma$ cross section, agreeing with the
L3 data and disagreeing the Opal results.  We stress that the $\gamma
p$ and the $\gamma\gamma$ experimental data are consistent with both
$P_{\rm had}$ being energy independent and the factorization theorem
of \eq{fact}.

We show that VMD, combined with quark counting, fits all available
`elastic' $\gamma p$ data. It even predicts correctly the phase of the
forward scattering amplitude for true Compton scattering, $\gamma
+p\rightarrow\gamma + p$. Our theory allows us to calculate that the
three light vector meson s, $\rho$, $\omega$ and $\phi$, account for
about 60\% of the total $\gamma p$ cross section.

Finally, there are also dynamical consequences of our phenomenological
model. We see that the gluons are carried along with the quarks, since
when we link our 2/3 factors exclusively to the quark composition of
our eikonal, we get strong disagreement with the measured nuclear
slope parameters $B$ for $\gamma p$ `elastic' interactions, as well as
disagreement between the $\rho$ values that we predict for `elastic'
scattering (such as $\gamma + p\rightarrow\rho^0+p$) and the
dispersion relation calculation\cite{gilman} for Compton scattering on
a proton, $\gamma +p\rightarrow\gamma+p$.


\newpage

\appendix
\setcounter{equation}{0} 
\renewcommand{\theequation}{\Alph{section}\arabic{equation}}
%
\section{Eikonal Formulation\label{app:eikonal}}
In order to ensure unitarity, we utilize an eikonal formalism,
evaluating the eikonal in the two-dimensional transverse impact
parameter space $\vec{b}$.  We introduce two eikonals, $\chi^{\rm
even}(b,s)$ and $\chi^{\rm odd}(b,s)$, even and odd under crossing,
respectively (where the proton labora tory energy ${\rm E}\rightarrow
-{\rm E}$), which are both complex and real analytic. In terms of the
even and odd eikonals, the eikonals we require for $pp$ and $\bar p p$
scattering are given by
\begin{eqnarray}
\chi_{p p}(b,s)&  =&
\chi^{\rm even}(b,s)-\chi^{\rm odd}(b,s)\nonumber  \\
 \chi_{\bar p p}(b,s)& = &
\chi^{\rm even}(b,s)+ \chi^{\rm odd}(b,s)\label{ppandpbarp}.
\end{eqnarray}

This work largely follows the procedures and conventions used by Block
and Cahn\cite{bc}.  In terms of the c.m. scattering amplitude $f_{\rm
c.m.}$, the c.m. differential elastic scattering cross section
$\frac{d\sigma}{d\Omega_{\rm c.m.}}(\theta)$, the invariant
differential elastic scattering distribution $\frac{d\sigma}{dt}$ and
the total cross section $\sigma_{\rm tot}$ are given by
\begin{eqnarray}
\frac{d\sigma}{d\Omega_{\rm c.m.}}(\theta)&=&\left|f_{\rm c.m.}\right|^2,\\
\frac{d\sigma}{dt}&=&\frac{\pi}{k^2}\left|f_{\rm c.m.}\right|^2,\label{dsdt0}\\
\sigma_{\rm tot}&=&\frac{4\pi}{k}{\rm Im}f_{\rm c.m.}(\theta=0),\label{optical}
\end{eqnarray}
where $k$ is the c.m. system momentum, $\theta$ is the c.m. system
scattering angle and $t=-2k^2(1-\cos \theta)$ is the invariant
four-momentum transfer.  Let $a(b,s)$ be the scattering amplitude in
impact parameter space. The c.m. scattering amplitude\cite{bc} is
given by
\begin{equation}
f_{\rm c.m.}(s,t)=\frac{k}{\pi}\int \,e^{i\vec q \cdot \vec
b}a(b,s)\,d^2\vec{b} ,\label{fcms}
\end{equation}
where $d^2\vec b=2\pi b\,db$, and $\vec{q}$ is a two-dimensional
vector in impact parameter space $\vec{b}$ such that $q^2=-t$.
Let the eikonal $\chi(b,s)$ be {\em complex}, such that
\be
\chi(b,s)=\chir (b,s)+ i\chii (b,s).\label{chi}
\ee
We define our eikonal $\chi(b,s)$ so that $a(b,s)$, the (complex)
scattering amplitude in impact parameter space $b$, is given by
\begin{eqnarray}
a(b,s)&=&\frac{i}{2}\left(1-e^{i\chi(b,s)}\right)\nonumber\\
&=&\frac{i}{2}\left(1-e^{-\chii(b,s)+i\chir(b,s)}
\right).\label{eik}
\end{eqnarray}
\subsection{Forward Scattering Parameters and Cross Sections
\label{app:forwardscattering}}
We now calculate the forward scattering parameters and various cross
sections, using the above eikonal formulation in impact parameter
space.
\subsubsection{Total Cross Section  $\sigma_{\rm tot}(s)$\label{app:sigtot}}
Using the optical theorem, the total cross section $\sigma_{\rm
tot}(s)$ is given by
\begin{eqnarray}
\sigma_{\rm tot}(s)&=&\frac{4\pi}{k}{\rm Im}\,f_{\rm c.m.}(s,0)\nonumber\\
&=&4\int \,{\rm Im}\,a(b,s)\,d^2\vec{b}\nonumber\\
&=&2\int\,\left[1-e^{-\chii
(b,s)}\cos(\chir(b,s))\right]\,d^2\vec{b},\label{sigtot}
\end{eqnarray}
where in \eq{sigtot}, we used \eq{fcms} evaluated in the forward
direction ($t=0$) and substituted \eq{eik} to evaluate the total cross
section in terms of the eikonal $\chi(b,s)$.
\subsubsection{Elastic Scattering Cross Section $\sigma_{\rm el}(s)$\label{app:sigel}}
>From \eq{dsdt0}, we can evaluate the elastic scattering cross section
$\sigma_{\rm el}(s)$ as
\begin{eqnarray}
\sigma_{\rm elastic}(s)&=&\frac{\pi}{k^2}\int \,|f_{\rm c.m.}(s,t)|^2\,dt\nonumber\\
&=&\frac{1}{k^2}\int \,|f_{\rm c.m.}(s,t)|^2\,d^2\vec{q}\nonumber\\
&=&\frac{1}{\pi^2}\int \int \int
e^{i\vec{q}\cdot(\vec{b}-\vec{b'})}a(b,s)a^*(b',s)
\,d^2\vec{q}\,d^2\vec{b}\,d^2\vec{b'}\nonumber\\
&=&\frac{(2\pi)^2}{\pi^2}\int \int
a(b,s)a^*(b',s)\delta^2(\vec{b}-\vec{b'})\,d^2\vec{q}\,d^2\vec{b}\nonumber\\
&=&4\int |a(b,s)|^2\,d^2\vec{b}\nonumber\\
&=&\int\left|1-e^{-\chii(b,s)+i\chir(b,s)}\right|^2\,d^2\vec{b}.
\label{sigel}
\end{eqnarray}
\subsubsection{Inelastic Cross Section $\sigma_{\rm in}(s)$\label{app:sigin}}
Thus, using
\eq{sigtot} and \eq{sigel},  
the inelastic cross section, defined as $\sigma_{\rm
tot}(s)-\sigma_{\rm elastic}(s)$, is given by
\be
\sigma_{\rm inelastic}(s)=\int\,\left\{1-e^{-2\chii(b,s)}\right\}\,d^2\vec{b}.
\label{sigin}
\ee
\subsubsection{Differential Elastic Scattering Cross Section $\frac{d\sigma}{dt}(s,t)$ 
\label{dsdt}}
A convenient way of calculating the differential elastic scattering
cross section of \eq{dsdt} is to note that an alternative way to write
\eq{fcms} is to introduce an integral representation of $J_0$ (see
eq. (9.1.18), ref. \cite{abrom&stegun}),
\be 
J_0(z)=\frac{1}{2\pi}\int_0^{2\pi}\, e^{iz\cos\phi}\,d\phi.
\ee
We can then rewrite \eq{fcms} as
\be
f_{\rm c.m.}(s,t)=2k\int_0^\infty\,J_0(qb)\,a(b,s)b\,db=\frac{k}{\pi}
\int\,J_0(qb)\,a(b,s)\,d^2\vec{b}.\label{newfcms}
\ee
Finally, using \eq{newfcms} and \eq{dsdt0}, we now write the
differential scattering cross section as
\be
\frac{d\sigma}{dt}(s,t)=\frac{1}{4\pi}\left|\int J_0(qb)(1-e^{-\chii(b,s)+i\chir(b,s)})\,d^2\vec{b}\,\right|^2,
\label{dsdt2}
\ee
a more convenient computational form.
\subsubsection{$\rho(s)$\label{app:rho}}
To calculate $\rho$, the ratio of the real to the imaginary portion of
the forward nuclear scattering amplitude, we write
\begin{eqnarray}
\rho(s)&=&\frac{{\rm Re}f_{\rm c.m.}(s,0)}{{\rm Im}f_{\rm c.m.}(s,0)}\nonumber\\
&=&\frac{{\rm Re}\left\{i(\int
1-e^{-\chii(b,s)+i\chir(b,s)})\,d^2\vec{b}\right\}} {{\rm
Im}\left\{i(\int
(1-e^{-\chii(b,s)+i\chir(b,s)})\,d^2\vec{b}\right\}}.\label{rho}
\end{eqnarray}
\subsubsection{Nuclear Slope Parameter $B$\label{app:B}}
The nuclear slope parameter $B$ is defined as
\be
B(s)=\frac{d}{dt}\left[\ln
\frac{d\sigma}{dt}(s,t)\right]_{t=0}.\label{B}
\ee
Beginning with \eq{fcms},
\be
f_{\rm c.m.}(s,t)\propto \int \,e^{i\vec q \cdot \vec
b}a(b,s)\,d^2\vec{b},
\ee
we expand the exponential about $q=0$ to get
\be
f_{\rm c.m.}(s,t)\propto \int [1+i\vec{q}\cdot\vec{b}
-\frac{1}{2}(\vec{q}\cdot\vec{b})^2+\cdots ]a(b,s)\,d^2\vec{b}
\ee
With this expansion and the definition of $B$ in \eq{B}, we can
eventually write the general expression for $B$ as
\be
B=\frac{{\rm Re}\left\{\int_0^\infty\,db\,b\,a(b,s)\int_0^\infty \,db
\,b^3a^*(b,s)\right\}}
{2\left|\int_0^\infty\,db\,b\,a(b,s)\right|^2}.\label{trueB}
\ee
If the phase of $a(b,s)$ is independent of b (this is the case when we
either have a {\em factorizable} eikonal or an eikonal with a constant
phase),
\eq{trueB} reduces to the more tractable expression 
\be
B=\frac{\int_0^\infty\,db\,b^3a(b,s)}{2\int_0^\infty\,db\,b\,a(b,s)}=
\frac{\int\,b^2a(b,s)\,d^2\vec{b}}{2\int\,a(b,s)\,d^2\vec{b}}.\label{Bsimple}
\ee
We note from \eq{Bsimple} that $B$ measures the size of the proton,
{\em i.e.,} $B$ is one-half the average value of the square of the
impact parameter $b$, weighted by $a(b,s).$ Again, introducing the
eikonal into \eq{Bsimple}, we find
\be
B=\frac{1}{2}\frac{\int\left(1-e^{-\chii(b,s)+i\chir(b,s)}\right)\,b^2\,\,d^2\vec{b}}
{\int\left(1-e^{-\chii(b,s)+i\chir(b,s)}\right)\,\,d^2\vec{b}},\label{Bfinal}
\ee
which we use to compute $B$.
\subsubsection{The Curvature Parameter $C$\label{app:curvature}}
The $t$ dependence of the elastic differential cross section is
described at small $|t|$ as
\begin{eqnarray}
 \frac{d\sigma}{dt}(s,t)&
 =&\left|\frac{d\sigma}{dt}(s,t)\right|_{t=0}e^{Bt+Ct^2+ \cdots}
 \nonumber\\ &= &
 \left|\frac{d\sigma}{dt}(s,t)\right|_{t=0}\left[1+Bt+\left(
\frac{B^2}{2}+C\right)t^2+\cdots\right].\label{curvature} 
\end{eqnarray}
We state, without proof, that the curvature parameter $C$ is
given\cite{bc} by
\begin{eqnarray}
C(s)&=&\frac{1}{32}\frac{\int
\left(1-e^{-\chii(b,s)+i\chir(b,s)}\right)b^4 \,d^2\vec{b}}{\int
\left(1-e^{-\chii(b,s)+i\chir(b,s)}\right)\,d^2\vec{b}}\nonumber\\
&&\qquad-\frac{1}{16}\left[\frac{\int
\left(1-e^{-\chii(b,s)+i\chir(b,s)}\right)b^2 \,d^2\vec{b}}{\int
\left(1-e^{-\chii(b,s)+i\chir(b,s)}\right)\,d^2\vec{b}}\right]^2,\label{calculateC}
\end{eqnarray}
where it was again assumed that the phase of the eikonal is
independent of $b$ (see eq. (4.45) of ref. \cite{bc}).  From
\eq{calculateC}, we see that the curvature $C$ can be positive,
negative or zero, whereas the nuclear slope parameter, from
\eq{Bfinal}, must be positive.
\setcounter{equation}{0}
\section{QCD-Inspired Eikonal\label{app:QCDeikonal}}
\subsection{Even Eikonal\label{app:QCDeven}}
The even QCD-Inspired eikonal $\chi_{\rm even}$ is given by the sum of
three contributions, glue-glue, quark-glue and quark-quark, which are
individually factorizable into a product of a cross section $\sigma
(s)$ times an impact parameter space distribution function
$W(b\,;\mu)$, {\em i.e.,}:
\begin{eqnarray}
 \chi^{\rm even}(s,b)& = &\chi_{\rm gg}(s,b)+\chi_{\rm
 qg}(s,b)+\chi_{\rm qq}(s,b)\nonumber\\ &=&i\left[\sigma_{\rm
 gg}(s)W(b\,;\mu_{\rm gg})+\sigma_{\rm qg}(s)W(b\,;\sqrt{\mu_{\rm
 qq}\mu_{\rm gg}})+\sigma_{\rm qq}(s)W(b\,;\mu_{\rm
 qq})\right],\label{chieven}
\end{eqnarray}
where the impact parameter space distribution function
\begin{equation}
W(b\,;\mu)=\frac{\mu^2}{96\pi}(\mu b)^3K_3(\mu b)\label{W}
\end{equation}
is normalized so that
\begin{equation}
\int W(b\,;\mu)d^2 \vec{b}=1. 
\end{equation}
Hence, the $\sigma$'s in \eq{chieven} have the dimensions of a cross
section.

The factor $i$ is inserted in \eq{chieven} since the high energy
eikonal is largely imaginary (the $\rho$ value for nucleon-nucleon
scattering is rather small).

As a consequence of both factorization and the normalization chosen
for the $W(b\,;\mu)$, it should be noted that
\be
\int \chi^{\rm even}(s,b)\, d^2\vec b=i\left[\sigma_{\rm gg}(s)+
\sigma_{\rm qg}(s)+\sigma_{\rm qq}(s)\right], \label{integrateeven}
\ee
so that, using \eq{sigtot} for {\em small} $\chi$,
\be
\sigma_{\rm tot}^{\rm even}(s)=2\,{\rm Im}\left\{ i\left[
\sigma_{\rm gg}(s)+\sigma_{\rm qg}(s)+\sigma_{\rm qq}(s)\right]\right\}
\label{smallevensigma}.
\ee

In \eq{chieven}, the inverse sizes (in impact parameter space)
$\mu_{\rm gg}$ and $\mu_{\rm gg}$ are to be fit by experiment, whereas
the quark-gluon inverse size is taken as $\sqrt{\mu_{\rm qq}\mu_{\rm
gg}}$.%
\subsubsection{The $\sigma_{\rm gg}$ Contribution\label{app:siggg}}
Modeling the glue-glue interaction after QCD, we write the cross
section $\sigma_{\rm gg}(s)$ in \eq {chieven} as
\begin{equation}
\sigma_{\rm gg}(s)=C_{\rm gg}N_{\rm g}^2\int\Sigma_{\rm gg}
\theta(\hat s-m_0^2)F_{\rm gg}\left ( x_1x_2=
\frac{\hat s}{s}\right )\,d\left (\frac{\hat s}{s}\right )\label{sigggqcd},
\end{equation}
where
\begin{equation}
\Sigma_{\rm gg}=\frac{9\pi \alpha_s^2}{m_0^2}.
\end{equation}
The normalization constant $C_{\rm gg}$ and the threshold $m_0$ are to
be fitted by experiment (in practice, the threshold is taken as $m_0 =
0.6$~GeV and the strong coupling constant $\alpha_s$ is fixed at
0.5). The constant $N_{\rm g}$ in \eq{sigggqcd} is given by $N_{\rm
g}=\frac{3}{2}\frac{(
5-\epsilon)(4-\epsilon)(3-\epsilon)(2-\epsilon)(1-\epsilon)}{5!}$.
Using the gluon structure function
\begin{equation}
f_{\rm g}(x)=3\frac{(1-x)^5}{x^{1+\epsilon}},\label{fgepsilon}
\end{equation} 
we can now write the function $F_{\rm gg}$ in \eq{sigggqcd} as
\begin{equation}
F_{\rm gg}=\int\int f_{\rm g}(x_1)f_{\rm g}(x_2)\delta
(x_1x_2=\tau)\,dx_1\,dx_2.
\end{equation}
After carrying out the integrations, we can now explicitly express
$\sigma_{\rm gg}(s)$ as a function of $s$. The parameter $\epsilon$ in
\eq{fgepsilon} is to be fitted by experiment (in practice, we fix it
at 0.05).
\paragraph{High Energy Behavior of $\sigma_{\rm gg}(s)$---the Froissart Bound
\label{app:highenergysigmagg}}
{\ }\newline 
{\ }\newline 
We note that the high energy
behavior of $\sigma_{\rm gg}(s)$ is controlled by
\begin{eqnarray}
\lim_{s \rightarrow\infty}\mbox{ }\int^1_{m^2_0/s} d\tau F_{gg}(\tau)&\sim&
\int^1_{m^2_0/s} d\tau\frac{{}-\log\tau}{\tau^{1+\epsilon}}\nonumber\\
&&\mbox{}\nonumber \\
&\sim&\left(\frac{s}{m^2_0}\right)^{\epsilon}\log
\left(\frac{s}{m^2_0}\right),
\end{eqnarray}
where $\epsilon >0$.  The cut-off impact parameter $b_c$ is given by
\begin{equation}
cW_{gg}(b_c;\mu_{gg})s^{\epsilon}\log(s)\sim 1,\label{about1}
\end{equation}
where $c$ is a constant.  For large values of $\mu b$, we can now
write \eq{about1} as
\begin{equation}
c'(\mu_{gg}b_c)^{3/2}e^{-\mu_{gg}b_c}s^{\epsilon}\log(s)\sim 1
\end{equation}
with $c'$ another constant, and therefore,
\begin{equation}
b_c=\frac{\epsilon}{\mu_{gg}} \log \frac{s}{s_0}+O\left (\log\log
\frac{s}{s'_0}\right).
\end{equation}
We reproduce the Froissart bound from QCD arguments,
\begin{equation}
\sigma_{tot}=2\pi \left(\frac{\epsilon}{\mu_{gg}}\right)^2\log^2
\frac{s}{s_0},\label{Froissart}
\end{equation}
as we go to very high energies, as long as $\epsilon>0.$ The usual
Froissart bound coefficient of the $\log^2\frac{s}{s_0}$ term,
$1/m^2_{\pi}=20$ mb, is now replaced by
$\left(\epsilon/\mu_{gg}\right)^2\sim 0.002$ mb. Note that $\mu_{gg}$
controls the size of the area occupied by the gluons inside the
nucleon.
\paragraph{Evaluation of the $\sigma_{\rm gg}$ Contribution \label{app:sigggevaluation}}
{\ }\newline 
{\ }\newline 
In the following, we set the
matrices $a(0)=-a(5)=-411/10$, $a(1)=-a(4)=-975/2,$ $a(2)=-a(3)=-600$
and $b(0)=b(5)=-9,$ $b(1)=b(4)=-225,$ $b(2)=b(3)=-900$.  The result is
\begin{eqnarray}
\sigma_{\rm gg}(s)&=&C_{\rm gg}\Sigma_{\rm gg}N_{\rm g}^2 
\int_{\tau_0}^{1}{F_{\rm gg}\,d\tau}\nonumber \\
&=&C_{\rm gg}\Sigma_{\rm gg}N_{\rm g}^2 \times \nonumber\\
&&\sum_{i=0}^{5}
\left\{
\frac{a(i)-\frac{b(i)}{i-\epsilon}}{i-\epsilon}
-\tau_0^{i-\epsilon}
\left(
\frac{a(i)-\frac{b(i)}{i-\epsilon}}{i-\epsilon}+\frac{b(i)}{i-\epsilon}\log (\tau_0)
\right)
\right\}
\nonumber\\	
&=&C_{\rm gg}\Sigma_{\rm gg}N_{\rm g}^2 \times\nonumber\\ &
&\!\!\!\!\!\!\!\!\left\{\quad \frac{
\frac{411}{10}+\frac{9}{\epsilon}}{\epsilon}
\ \quad -\tau_0^{-\epsilon}
\left(\frac{ \frac{411}{10}+\frac{9}{\epsilon}}{\epsilon}+
\frac {9\log(\tau_0)}{\epsilon}\right)\right.\nonumber \\
 &+ & \frac{ \frac{-975}{2}+\frac{225}{1-\epsilon}}{1-\epsilon}
 -\tau_0^{1-\epsilon}
\left(\frac{ \frac{-975}{2}+\frac{225}{1-\epsilon}}{1-\epsilon}-
\frac {225\log(\tau_0)}{1-\epsilon}\right)\nonumber  \\
 & + & \frac{-600+\frac{900}{2-\epsilon}}{2-\epsilon}
 -\tau_0^{2-\epsilon}
\left(\frac{ -600+\frac{900}{2-\epsilon}}{2-\epsilon}-
\frac {900\log(\tau_0)}{2-\epsilon}\right)\nonumber  \\
 & + & \frac{600+\frac{900}{3-\epsilon}}{3-\epsilon}
 -\tau_0^{3-\epsilon}
\left(\frac{600+\frac{900}{3-\epsilon}}{3-\epsilon}-
\frac {900\log(\tau_0)}{3-\epsilon}\right)\nonumber  \\
& + & \frac{ \frac{975}{2}+\frac{225}{4-\epsilon}}{4-\epsilon}
-\tau_0^{4-\epsilon}
\left(\frac{ \frac{975}{2}+\frac{225}{4-\epsilon}}{4-\epsilon}-
\frac {225\log(\tau_0)}{4-\epsilon}\right)\nonumber \\
 &+& \left.\frac{ \frac{411}{10}+\frac{9}{5-\epsilon}}{5-\epsilon}
 -\tau_0^{5-\epsilon}
\left(\frac{ \frac{411}{10}+\frac{9}{5-\epsilon}}{5-\epsilon}-
\frac {9\log(\tau_0)}{5-\epsilon}\right)\right\}
,\qquad {\rm where\ }\tau_0= \frac{{m_0}^2}{s}.
\label{Fggintegrated}
\end{eqnarray}

We note that we must fit the following 3 constants in order to specify
$\sigma_{\rm gg}$:
\begin{enumerate}
\item the normalization constant $C_{\rm gg}$.
\item the threshold mass $m_0$.
\item $\epsilon$, the parameter in the gluon structure function which determines 
the behavior at low x  ($\propto 1/x ^{1+\epsilon}$).
\end{enumerate}
\subsection{The $\sigma_{\rm qq}$ Contribution\label{app:sigqq}}
If we use the toy structure function
\begin{equation}
f_{\rm q}(x)=\frac{(1-x)^3}{\sqrt x},\label{fq}
\end{equation}%
we can write
\begin{eqnarray}
 \sigma_{\rm qq}(s)& \propto &
 \frac{m_0}{\sqrt{s}}\log\frac{s}{s_0}+{\cal P}\left(\frac{m_0}{\sqrt
 s}\right) \nonumber\\ & \approx & {\rm constant}+ \frac{m_0}{\sqrt
 s},
\end{eqnarray}
where $\cal P$ is a polynomial in $m_0/{\sqrt s}$.

Thus, we approximate the quark-quark term by
\begin{equation}
 \sigma_{\rm qq}(s)=\Sigma_{\rm gg} \left( C + C_{\rm Regge\ even}
 \frac {m_0}{\sqrt s}\right ), \label{sigmaqq} \end{equation} where
 $C$ and $C_{\rm Regge\ even}$ are constants. Thus, $\sigma_{\rm
 qq}(s)$ simulates quark-quark interactions with a constant cross
 section plus a Regge-even falling cross section.

We must fit the following 2 constants in order to specify $\sigma_{\rm
qq}$:
\begin{enumerate}
\item the normalization constant $C$.
\item the normalization constant $C_{\rm Regge\ even}$.
\end{enumerate}
\subsection{The $\sigma_{\rm qg}$ Contribution\label{app:sigqg}}
If we use the toy structure function
\begin{equation}
f_{\rm g}(x)=\frac{(1-x)^5}{x},\label{fg}
\end{equation}
and the toy structure function $f_{\rm q}(x)$ of \eq{fq} we can write
\begin{eqnarray}
 \sigma_{\rm qg}(s)& \propto & C''\log \frac{s}{s_0}+C'{\cal
 P'}\left(\frac{m_0}{\sqrt s}\right) \nonumber\\ & \approx & C''\log
 \frac{s}{s_0}+ C',
\end{eqnarray}
 where $C'$ and $C''$ are constants and $\cal P'$ is a polynomial in
 $m_0/{\sqrt s}$.

Thus, if we absorb the constant piece $C'$ into the quark-quark term,
we can approximate the quark-gluon term by
\begin{equation}
 \sigma_{\rm qg}(s)=\Sigma_{\rm gg} C_{\rm qg\ log}\log\frac{s}{s_0},
 \label{sigmaqg} \end{equation} where $C_{\rm qg\ log}$ is a constant.
 Hence, we attempt to simulate diffraction with the logarithmic term
 $\sigma_{\rm qg}(s)$.

We must fit the following 2 constants in order to specify $\sigma_{\rm
qg}$:
\begin{enumerate}
\item the normalization constant $C_{\rm qg\ log}$.
\item $s_0$, the square of the energy scale in the log term of \eq{sigmaqg}.
\end{enumerate}

\subsubsection{Making the Even Contribution Analytic\label{app:evencontribution}}
The total even contribution, which is not yet analytic, can be written
as the sum of equations \ref{Fggintegrated}, \ref{sigmaqq} and
\ref{sigmaqg}, {\em i.e.,}
\begin{eqnarray}
\chi_{\rm even}&=&i\left\{
\vphantom{
\left.\Sigma_{\rm gg}\left [ \left( C + C_{\rm Regge\  even} 
\frac {m_0}{\sqrt s}\right )W(b\,;\mu_{\rm qq})+ C_{\rm qg\ log}
\log\frac{s}{s_0}W(b\,;\sqrt{\mu_{qq}\mu_{gg}})\right]\right\}
}%
\sigma_{\rm gg}(s)W(b\,;\mu_{\rm gg})\right.\nonumber\\
&&\quad+\Sigma_{\rm gg} \left( C + C_{\rm Regge\ even} \frac
{m_0}{\sqrt s}\right )W(b\,;\mu_{\rm qq})\nonumber\\ &&\,\,\,\quad +
\left. \Sigma_{\rm gg} C_{\rm qg\
log}\log\frac{s}{s_0}W(b\,;\sqrt{\mu_{qq}\mu_{gg}})\right\}.\label{finaleven}
\end{eqnarray}
For large $s$, the {\rm even} amplitude in \eq{finaleven} can be made
analytic by the substitution (see the table on p. 580 of reference
\cite{bc}, along with reference \cite{eden})
\[s\rightarrow se^{-i\pi/2}.\] 
Thus, we finally rewrite the even contribution of \eq{finaleven},
which is now {\em analytic}, as
\begin{eqnarray}
\chi_{\rm even}&=&i\left\{
\vphantom{
\left.\Sigma_{\rm gg}\left [ \left( C + C_{\rm Regge\  even} \frac {m_0}{\sqrt s}
\right )W(b\,;\mu_{\rm qq})+ C_{\rm qg\ log}\log\frac{s}{s_0}W(b\,;\sqrt{\mu_{qq}
\mu_{gg}})\right]\right\}
}%
\sigma_{\rm gg}(se^{-i\pi/2})W(b\,;\mu_{\rm gg})\right.\nonumber\\
&&\quad+\Sigma_{\rm gg}\left( C + C_{\rm Regge\ even} \frac
{m_0}{\sqrt s}e^{i\pi/4}\right )W(b\,;\mu_{\rm qq})\nonumber\\
&&\,\,\,\quad + \left. \Sigma_{\rm gg}C_{\rm qg\log}
\left(\log\frac{s}{s_0}-i\frac{\pi}{2}\right)W(b\,;\sqrt{\mu_{qq}\mu_{gg}})
\right\}.\label{evenanalytic}
\end{eqnarray}

To determine the impact parameter profiles in $b$ space, we also must
fit the mass parameters $\mu_{\rm gg}$ and $\mu_{\rm qq}$ to the
data. We find masses $\mu_{\rm gg}\approx 0.73$ GeV and $\mu_{\rm
qq}\approx 0.89$ GeV.
\subsection{The Odd Eikonal\label{app:oddeikonal}}
It can be shown that a high energy analytic {\em odd} amplitude (for
its structure in $s$, see eq. (5.5b) of reference \cite{bc}, with
$\alpha =0.5$) that fits the data is given by
\begin{eqnarray}
\chii^{\rm odd}(b,s)&=&-\sigma_{\rm odd}\,W(b;\mu_{\rm odd})\nonumber\\
&=&-C_{\rm odd}\Sigma_{\rm
gg}\frac{m_0}{\sqrt{s}}e^{i\pi/4}\,W(b;\mu_{\rm
odd}),\label{oddanalytic}
\end{eqnarray}
with
\be
W(b,\mu_{\rm odd})=\frac{\mu_{\rm odd}^2}{96\pi}(\mu_{\rm odd}
b)^3\,K_3(\mu_{\rm odd} b)\label{Woddnormalization},
\ee
normalized so that
\begin{equation}
\int W(b\,;\mu)d^2 \vec{b}=1. \label{oddWintegral}
\end{equation}
Hence, the $\sigma_{\rm odd}$ in \eq{oddanalytic} has the dimensions
of a cross section.

In order for $C_{\rm odd}$ to be positive, a minus sign has been
inserted in \eq{oddanalytic}.

With the normalization (\eq{Woddnormalization} and \eq{oddWintegral})
chosen for $W(b,\mu_{\rm odd})$, we see that
\be
\int \chi^{\rm odd}(s,b)\, d^2\vec b=\sigma_{\rm odd}(s)\label{integrateodd},
\ee
so that, using \eq{sigtot} for {\em small} $\chi$,
\be
\sigma_{\rm tot}^{\rm odd}(s)=2\,{\rm Im}\,\sigma_{\rm odd}(s)\label{smalloddsigma}.
\ee

In order to determine the cross section $\sigma_{\rm odd}$, we must
fit the normalization constant $C_{\rm odd}$. To determine the impact
parameter profile in $b$ space, we also must fit the mass parameter
$\mu_{\rm odd}$ to the data. We find a mass $\mu_{\rm odd}\approx
0.53$ GeV.

We again reiterate that the odd eikonal, which we see (from
\eq{oddanalytic}) vanishes like $\frac{1}{\sqrt s}$, accounts for the
{\em difference} between $pp$ and $\bar p p$. Thus, at high energies,
the odd term vanishes, and we can neglect the difference between $pp$
and $\bar p p$ interactions .
%
%

%
%
\end{document}